\documentstyle[aps,amsfonts,twocolumn]{revtex}
\def\bbbk{{\rm I\!K}}
\newcommand{\sign}{\mathop{\rm sign}\nolimits}
\newcommand{\la}{\langle}
\newcommand{\ra}{\rangle}
\newcommand{\g}{\hbox{\sl g}}
\newcommand{\D}{\delta}
\newcommand{\dd}{\text{d}}
\newcommand{\B}[1]{\hbox{$\protect\bbox{{\Bbb #1}}$}}
\newcommand{\BT}[1]{\hbox{$\widetilde{\protect\bbox{{\Bbb #1}}}$}}
\newcommand{\XT}{\hbox{$\widetilde{X}$}}
\newcommand{\Epsilon}{{\cal E}}
\newcommand{\CC}{C}
\newlength{\LENs}
\newlength{\LENn}
\newlength{\LENl}
\newlength{\LENL}
\def\LENset{\ifpreprintsty \LENs=2.8mm \LENn=3.0mm \LENl=4.5mm \LENL=5mm%
\else \LENs=1.16mm \LENn=1.9mm \LENl=3.4mm \LENL=3.8mm \fi}
\LENset
\newcommand{\m}[1]{\makebox[\LENs][c]{\raisebox{0.3mm}{-}$#1$}}
\newcommand{\p}[1]{\makebox[\LENs][c]{$#1$}}

\newcommand{\ml}[1]{\makebox[\LENn][c]{\raisebox{0.3mm}{-}$#1$}}
\newcommand{\Pl}[1]{\makebox[\LENn][c]{$#1$}}

\newcommand{\mL}[1]{\makebox[\LENl][c]{-$#1$}}
\newcommand{\pL}[1]{\makebox[\LENl][c]{$#1$}}
\newcommand{\ii}[1]{\makebox[0mm][c]{{\footnotesize $\Ss #1$}}}
\newcommand{\ib}[1]{\hbox{\footnotesize $\Ss#1$}}
\newcommand{\Ss}{\scriptstyle}
\newcommand{\Tld}[1]{\hbox{$\vphantom{#1}\smash{\widetilde{#1}}$}}
\newcommand{\tld}[1]{\hbox{$\vphantom{#1}\smash{\tilde{#1}}$}}
\newcommand{\otm}{\hbox{$\otimes$}}
\newcommand{\eref}[1]{\hbox{(\ref{#1})}}
\newcommand{\vsp}[1]{\noalign{\vspace*{#1}}}
\newcommand{\mass}{\text{m}}
\newcommand{\mc}{\frac{\mass\Cdot c}{2}}
\newcommand{\NN}{\nonumber}
\newcommand{\bpsi}{\bbox{\psi}}
\newcommand{\Cdot}{\,}

\begin{document}

\title{Generalized equation of relativistic quantum mechanics}

\author{A. A. Ketsaris}
\address{19-1-83, ul. Krasniy Kazanetz, Moscow 111395, Russian Federation}
\date{\today}
\maketitle

\begin{abstract}

We develop a new concept of quantum mechanics which is based on a generalized
space-time and on an action vector space similar to it.  Both spaces are
provided by algebraic properties.  This allows {\it to calculate} the Dirac
matrixes and {\it to derive} quantum mechanics equations from structure
equations of the specified algebras. A new interpretation of the wave
function is given as differential of the action vector.  A generalization of
the Dirac equation for 8-component wave function is derived. It is
interpreted as the equation for two leptons of the same generation. A
procedure of the approximate description of free leptons is formulated. The
generalized equation of quantum mechanics is reduced to the Dirac, Pauli and
Schr\"odinger equations by the sequential use of this procedure.  We explain
the existence of three lepton generations.

\end{abstract}



\section{Introduction}

Until the present time the sharp distinctions between the classical and
quantum mechanics principles gives no way of describing macro- and
micro-Universes in the uniform context. Let us remind of some of them.

In the classical mechanics, some operator can be put into correspondence with
any physical observable.  For example, the operator of derivation by
coordinate corresponds to the impulse.  The classical operators work in the
space of fundamental {\it scalar} quantity, the action $S$, over the {\it
reals}.  Thus the physical observable is identified with a result of
application of the operator to the function $S$.  There are no limitations on
the physical observable except for those imposed by the equation of
motion.

In the quantum mechanics, some operator corresponds also to any physical
observable.  The quantum operators work in the space of fundamental {\it
vector} quantity, wave function $\psi$, over the {\it complex}.
Thus the physical observable is identified not with a result
of application of operator to function $\psi$ but with the eigenvalues of
this operator.  These eigenvalues limit possible values of physical
observable (apart from restrictions imposed by the quantum mechanics
equations on the wave function).

This paper is aimed at finding the uniform basis both for classical and
for quantum mechanics. We follow the concept that includes two essential
generalizations. The first one concerns the space-time, the second one
concerns the action.

Our special purpose is to derive the relativistic quantum mechanics
equations. It is assumed that such a derivation will lead to the
understanding wave function through properties of generalized space-time and
generalized action and will also introduce a new quality to the equations
themselves.

The following propositions are used as the basis for our investigation:
\begin {enumerate}
\item
The space-time, $X$, is generalized up to $\B{X}$, space of all contravariant
tensors over $X$. The vector space $\B{X}$ supplemented by the {\it vector
multiplication} rule is algebra.  The space $\B{X}$ is defined as a {\it
generalized space-time}.  It is assumed that the generalized
space-time is a space of elementary particles.
The Clifford algebra, $\B{C}$, is further selected from $\B{X}$.
Its space is considered as a {\it space of leptons}.

\item
Apart from of the generalized space-time $\B{X}$, a generalized {\it
conjugate} space-time, $\BT{X}$, is introduced as a set of all covariant
tensors over $X$.  $\BT{X}$ is also algebra.  The conjugate space-time is
identified with a space of elementary antiparticles.  The space of the
conjugate Clifford algebra $\BT{C}$ selected from $\BT{X}$ is considered as a
{\it space of antileptons}.  For the algebra $\BT{C}$, as we show, the
approximate regular representation of basis vectors is given by the Pauli
and Dirac matrices.

\item
The action is considered as the {\it vector} quantity. The action vectors
form an algebra, $\B{S}$, {\it similar} to $\B{X}$. Moreover a space of {\it
conjugate} action vectors, $\BT{S}$, is introduced as similar to $\BT{X}$.
The spaces $\B{S}$ and $\BT{S}$ are related to elementary particles and
antiparticles. The Clifford algebras $\B{SC}$ and $\BT{SC}$, selected from
algebras $\B{S}$ and $\BT{S}$, are related with leptons and antileptons. They
are similar to the Clifford algebras $\B{C}$ and $\BT{C}$, respectively.

\item
The partial derivation of the multiplication rule in the algebras $\B{X}$ and
$\BT{X}$, $\B{S}$ and $\BT{S}$ produce specific differential relations
called {\it structure equations}.  The structure
equations for the Clifford algebras $\B{SC}$ and $\BT{SC}$ are reduced to the
generalized equations of relativistic quantum mechanics for leptons and
antileptons, respectively.  The relativistic quantum mechanics equations are
reduced to the Dirac equations with the simplified assumptions.

\end{enumerate}
\section{Generalization of space-time}
\subsection{Space-time}
\label{Conjugate_space}

Consider the space-time $X$ as a vector space over the reals,
$\bbbk$. A vector $x \in X$ can be expressed through basis vectors
\[
     x = e_i\Cdot {x}^i  \,,
\]
where $i=1,2,3,4$; $e_1$, $e_2$, $e_3$ are the basis vectors of geometric
space, $e_4$ is the basis vector of time, and $x^i \in \bbbk$ are vector
coordinates.

Let us introduce an operation inverse to multiplication of vector by number.
For each vector $a\in X$, there is a linear transformation
of vector $x\in X$ onto $\bbbk$, which is called {\it the scalar product}
of vectors $a$ and $x$ and is denoted by
\[
     \la a,x\ra \in \bbbk  \,.
\]
The vectors $a$ and $x$ are {\it orthogonal} to each other if
\[
     \la a,x\ra =0  \,.
\]
The scalar product of vector $x$ by itself defines its {\it length square}
\[
     \la x,x\ra = (x^1)^2 + (x^2)^2 + (x^3)^2 - (x^4)^2 \,.
\]
The {\it metric tensor} in $X$ is given by the scalar product of basis vectors:
{\tighten
\[
\raisebox{-2.0mm}{$\g_{ik} \equiv \la e_i,e_k\ra ={}$}
\hbox{\footnotesize
\tabcolsep=1mm
\renewcommand{\arraystretch}{0.95}
\begin{tabular}{c|cccc|c}
\multicolumn{1}{r}{}&\ib{1}&\ib{2}&\ib{3}&\multicolumn{1}{c}{\ib{4}}\\
\cline{2-5}
\ib{1}&\p 1 &   &   &   \\
\ib{2}&   &\p 1 &   &   \\
\ib{3}&   &   &\p 1 &   \\
\ib{4}&   &   &   &\m 1 & \,.\\
\cline{2-5}
\end{tabular}}
\]}
We omit hereafter zero matrix elements for convenience.
The {\it inverse} metric tensor $\g^{ik}$ is defined by condition
\[
     \g^{ik}\Cdot \g_{kl} = {\delta^i\!}_l \,.
\]

Let us introduce basis vectors $E^i\in X$ for which
\[
     \la E^i,x\ra =x^i  \,.
\]
The basis vectors $E^i$ are called {\it conjugate} with respect to basis
vectors $e_i$. For conjugate basis vectors
\[
     \la E^i, e_k\ra = {\delta^i\!}_k  \,.
\]
The conjugate basis vectors $E^i$ are connected to the basis vectors $e_i$ by a
relation:
\[
      E^i = e_k\Cdot \g^{ik}  \,.
\]
The {\it conjugate} vector
\[
     \tld{x} = x_i\Cdot E^i
\]
can be put into correspondence with the vector $x = e_i\Cdot x^i$.
Here $x_i = \delta_{ik}\Cdot x^k$ are coordinates of conjugate vector.
The scalar product of vector $x$ by conjugate vector $\tld{x}$ is
\[
     \la x,\tld{x}\ra = (x^1)^2 + (x^2)^2 + (x^3)^2 + (x^4)^2 \,.
\]
A set of conjugate vectors forms the vector space over the reals which will
be called {\it a conjugate space-time} and will be denoted by $\XT$.

\subsection{Generalized space-time}
\label{Gen-Space-Time}

Let us generalize the space-time $X$ up to the universal algebra of
contravariant tensors, $\B{X}$. For this purpose we introduce a
{\it tensor multiplication} $\otm$ for vectors $x_1\in X$ and consider, in
addition to vectors $x_1$, pairs $x_1\otm
x_2$, triples $x_1\otm x_2\otm x_3$
and, in the common case, a set of $\otm$-products of $p$ vectors,
$x_1\otm x_2\otm \ldots\otm x_p$, where $x_2,\ldots,x_p$ belong also to $X$.
A space of vectors
\[
     x_1\otm x_2\otm \ldots\otm x_p \,,
\]
will be denoted by $X^p$.
A vector $x\in X^p$ can be expressed using basis vectors
\[
     x = e_{i_p \ldots i_2i_1}\Cdot x^{i_1 i_2\ldots i_p} \,,
\]
where
\[
     e_{i_p \ldots i_2i_1} = e_{i_1}\otimes
     e_{i_2}\otimes\ldots \otimes e_{i_p} \,.
\]
Let us introduce a vector space
\[
     \B{X} = X^0 + X^1 +\ldots  +X^p + \ldots\,,
\]
where $X^0=\bbbk$, $X^1=X$. The space $\B{X}$ with the addition and
multiplication of vectors is {\it the universal algebra of contravariant
tensors} or {\it the universal contravariant algebra}.  The vector $x\in
\B{X}$ can be expressed through basis vectors
\[
       x = e_0\Cdot x^0 + e_{i_1}\Cdot x^{i_1} +
           e_{i_2 i_1}\Cdot x^{i_1 i_2} + \ldots +
           e_{i_n \ldots i_2i_1}\Cdot x^{i_1 i_2\ldots i_n} + \ldots
\]
Here the unit of reals $\bbbk$ is designated by $e_0$.
The vector space $\B{X}$ will be called {\it a generalized space-time}.

For basis vectors of $\B{X}$-space, the multiplication is given by
\[
     e_{i_{m+n}\ldots i_{m+2} i_{m+1}}\otimes e_{i_m\ldots i_2 i_1} =
     e_{i_{m+n}\ldots i_2 i_1}\,,
\]
or more commonly,
\[
     e_{i_m \ldots i_2i_1}\otimes e_{k_n \ldots k_2k_1} =
     e_{l_{m+n}\ldots l_2 l_1}\Cdot
     \delta_{i_m \ldots i_2 i_1 k_n \ldots k_2 k_1}^{l_1 l_2\ldots l_{m+n}}\,.
\]
An {\it upper generalized index}
\[
     I_m = i_1 i_2\ldots i_m\,
\]
and a {\it lower generalized index}
\[
     I_m = i_m \ldots i_2i_1\,.
\]
are introduced to make this expression more compact.
We can rewrite
\[
           e_{I_m} = e_{i_m \ldots i_2i_1}\,,\qquad
           x^{K_n} = x^{k_1 k_2\ldots k_n}\,.
\]
Then the rule of multiplication of basis
vectors in $\B{X}$-algebra takes the form
\begin{equation}
           e_{I_m}\otimes e_{K_n} = e_{L_{m+n}}\cdot
           {\delta^{L_{m+n}}\!}_{I_m K_n}\,.
\label{F01}
\end{equation}
Here the Kronecker deltas ${\delta^{L_{m+n}}\!}_{I_m K_n}$ can be
considered as {\it structure constants} of $\B{X}$-algebra.

Let us introduce an {\it upper collective index} $I$ running values
\[
       0, i_1, (i_1 i_2),\ldots, (i_1 i_2\ldots i_n),\ldots
\]
and a {\it lower collective index} $I$ running values
\[
       0, i_1, (i_2 i_1),\ldots, (i_n \ldots i_2 i_1),\ldots
\]
Then the vector $x\in \B{X}$ can be rewritten in a compact form:
\[
     x = e_I\cdot x^I\,.
\]

\subsection{Subspace of generalized space-time}

In this section we study how subspace and subalgebra are selected from
the generalized space-time.

In spite of the fact that the $\B{X}$-space is infinite-dimensional, we
introduce formally its dimensionality, $N$.
Let $q$ coordinates be expressed through $p=N-q$ the other coordinates
for any vector $x\in\B{X}$:
\[
      x^{I(q)} = {A^{I(q)}\!}_{K(p)}\cdot x^{K(p)} \,.
\]
Here the collective index $I(q)$ runs index values of dependent coordinates,
the collective index $K(p)$ runs index values of independent coordinates, and
${A^{I(q)}\!}_{K(p)}$ are connecting constants.
Then vector
\begin{eqnarray*}
      x &=&
      e_K\cdot x^K = e_{K(p)}\cdot x^{K(p)} + e_{I(q)}\cdot x^{I(q)} \\
        &=& \left(e_{K(p)} +
      e_{I(q)} \cdot {A^{I(q)}\!}_{K(p)}\right)\Cdot x^{K(p)} \\
\end{eqnarray*}
belongs to the subspace of $\B{X}$ with dimensionality $p$.
This subspace will be denoted by $\B{D}$. Basis vectors of $\B{D}$ are
\begin{eqnarray}
      \varepsilon_{K(p)}
      &=& e_{K(p)} + e_{I(q)}\cdot {A^{I(q)}\!}_{K(p)} \NN\\
      &=& e_{I(N)}\cdot {A^{I(N)}\!}_{K(p)} \,,
\label{F02}
\end{eqnarray}
where we have introduced the notation
\[
     {A^{I(N)}\!}_{K(p)} =
     \left\{
     \begin{array}{ll}
          {\delta^{I(p)}\!}_{K(p)}\,, &\hbox{for } I(N)=I(p)\,;\\
          {A^{I(q)}\!}_{K(p)}\,,      &\hbox{for } I(N)=I(q)\,.\\
     \end{array}
     \right.
\]

Hereafter we use the large latin letters for indices of basis vectors and of
coordinates in $\B{D}$-subspace as well as in $\B{X}$-space (for example,
$\varepsilon_K$ instead of $\varepsilon_{K(n)}$).  Thus it is meant
that the indices in $\B{D}$-subspace accept only $p$ values.

We shall now find the condition such that the subspace of $\B{X}$
is also {\it subalgebra} of $\B{X}$.  If $\B{D}$ is algebra with finite
dimensionality then the multiplication rule of basis vectors $\varepsilon_K$
can be written as
\begin{equation}
      \varepsilon_I\circ \varepsilon_K = \varepsilon_L\cdot {\CC^L\!}_{IK} \,.
\label{F03}
\end{equation}
Here ${\CC^L\!}_{IK}$ are the structure constants or the parastrophic
matrices of algebra $\B{D}$, the symbol "$\circ$" designates multiplication
unlike $\otimes$-multiplication used for $\B{X}$-algebra.  We  establish
connection between the multiplication rules in $\B{D}$ and in $\B{X}$.  We
rewrite the relation \eref{F02} as
\[
      \varepsilon_K = e_{K_n} \cdot {A^{K_n}\!}_K \,,
\]
the summation is over all basis vectors of the space $\B{X}$.
From it and from \eref{F01} we derive
\begin{eqnarray*}
      \varepsilon_I\circ \varepsilon_K
      &=& e_{I_m} \otimes e_{K_n}\cdot
      {A^{I_m}\!}_I\cdot {A^{K_n}\!}_K \\
      &=& e_{L_{m+n}}\cdot {\delta^{L_{m+n}}\!}_{I_m K_n}
      {A^{I_m}\!}_I\cdot {A^{K_n}\!}_K \\
      &=&
      e_{L_{m+n}}\cdot {A^{L_{m+n}}\!}_L\cdot {\CC^L\!}_{IK}\,.
\end{eqnarray*}
Hereof we obtain the relation between the structure
constants of subalgebra $\B{D}$ and the relation constants:
\[
      {\delta^{L_{m+n}}\!}_{I_m K_n}\cdot {A^{I_m}\!}_I\cdot {A^{K_n}\!}_K =
      {A^{L_{m+n}}\!}_L\cdot {\CC^L\!}_{IK}\,.
\]
This relation represents the condition when the subspace $\B{D}$ of
generalized space-time is the subalgebra of $\B{X}$.

In a specific case the multiplication \eref{F03} determines the scalar
product of basis vectors
\[
     \la \varepsilon_I,\varepsilon_K\ra=\varepsilon_0\Cdot {\CC^0\!}_{IK}
\]
and the metric tensor
\[
     \g_{IK}={\CC^0\!}_{IK}\,.
\]
Note also that
\[
     {\CC^L\!}_{I0}={\delta^L\!}_I\,,\qquad  {\CC^L\!}_{0K}={\delta^L\!}_K\,.
\]

$\B{D}$ is algebra {\it with division} if for each
vector $x\in\B{D}$ except a zero-vector,
there is the {\it inverse} vector $x^{-1}$ that satisfies the relation
\begin{equation}
     x\circ x^{-1} = \varepsilon_0\,.
\label{F04}
\end{equation}
Or in the coordinate form
\[
      \g_{IK}\cdot x^I\Cdot(x^{-1})^K = 1\,.
\]

\subsection{Regular representation of subalgebra of generalized space-time}

The universal contravariant algebra $\B{X}$ is associative. Therefore its
subalgebra $\B{D}$ is also associative. From the associativity of a
subalgebra $\B{D}$
\[
       (\varepsilon_N\circ \varepsilon_I)\circ \varepsilon_K =
       \varepsilon_N\circ (\varepsilon_I\circ \varepsilon_K)
\]
its regular representation follows.
Using \eref{F03} we obtain
\[
       (\varepsilon_L\circ \varepsilon_K)\Cdot {\CC^L\!}_{NI}
       = (\varepsilon_N\circ \varepsilon_L)\Cdot {\CC^L\!}_{IK} \,.
\]
Herefrom
\[
       {\CC^M\!}_{LK}\cdot {\CC^L\!}_{NI}  =
       {\CC^M\!}_{NL}\cdot {\CC^L\!}_{IK}\,.
\]
Comparing this expression with \eref{F03} we see that it is possible to
put parastrophic matrices ${C^L\!}_{NI}$ into a correspondence with the basis
vectors $\varepsilon_I$.  This correspondence is called {\it the regular (joined)
representation} of the $\B{D}$-algebra and is denoted as
\[
       \varepsilon_I \sim {\CC^L\!}_{NI} \,.
\]
The number $I$ of parastrophic matrix is the index of basis vector that can
be represented by this matrix.

\subsection{Generalized space-time of leptons}

Consider the {\it Clifford algebra} as subalgebra of universal contravariant
algebra $\B{X}$.  As we shall see later, this algebra holds a central position
in lepton physics. We shall define the Clifford algebra in two steps.

At first we define a {\it contracted algebra}, $\B{R}$, as subalgebra of
$\B{X}$ through the following conditions on coordinates of vectors:
\[
      x^{i_1 i_2\ldots i_p(k_1 k_1)i_{p+1}\ldots
      i_q(k_m k_m)i_{q+1}\ldots i_n}
      {}=
      x^{i_1 i_2\ldots i_n}\Cdot
      \prod\limits_{l=1}^m \g^{k_l k_l} \,.
\]
Here the $k$-indices in brackets are equal to each other, $m$ is the
number of pairs of such indices; any two neighbouring
$i$-indices are not equal to each other. Thus according to
\eref{F02} the vector space of algebra $\B{R}$ is built on the basis vectors
\begin{eqnarray*}
      &&\varepsilon_{i_n \ldots i_2 i_1} = e_{i_n \ldots i_2 i_1} \\
      &&\quad{}+
      \sum\limits_m
      e_{i_n\ldots i_{q+1}(k_m k_m)i_q\ldots
      i_{p+1}(k_1 k_1)i_p\ldots i_2 i_1}
      \Cdot \prod\limits_{l=1}^m \g^{k_l k_l} \,,
\end{eqnarray*}
non-containing identical neighbouring indices.
In particular,
\[
      \varepsilon_{i_2 i_1} = \varepsilon_{i_1} \circ \varepsilon_{i_2} =
      \left\{
      \begin{array}{ll}
          e_0 \Cdot \g_{i_1i_1},                & \hbox{for } i_1=i_2    \,;\\
          e_{i_1}\otimes e_{i_2} = e_{i_2 i_1}, & \hbox{for } i_1\ne i_2 \,.\\
      \end{array}
      \right.
\]
Thus $\varepsilon_{ii}={\CC_{ii}\!}^0 \Cdot e_0$ is the scalar product $\la
e_i, e_i\ra = ({e_i})^2$, and ${\CC_{ii}\!}^0$ is the metric tensor
$\g_{ii}$.  For example, if the $X$-space is one-dimensional and
$({e_1})^2=1$ then the $\B{R}$-space is constructed on the two basis vectors:
\begin{eqnarray*}
     \varepsilon_0 &=& e_0 + e_{11}  + e_{1111} + \ldots  +
     e_{\underbrace{\scriptstyle 11\ldots1}_{2k}} + \ldots \,,\\
     \varepsilon_1 &=& e_1 + e_{111}  + e_{11111} + \ldots  +
     e_{\underbrace{\scriptstyle 11\ldots1}_{2k+1}} + \ldots
\end{eqnarray*}

Now we define the Clifford algebra $\B{C}$ as subalgebra of $\B{R}$
by the following system of linear equations:
\[
     x^{\sigma(i_1i_2\ldots i_n)} = - x^{i_1i_2\ldots i_n} \,,
\]
where $\sigma$ is the permutation of any two neighbouring distinct indices.
For example,
\[
      x^{43142} = - x^{34142} = x^{31442} = x^{312}\Cdot\g^{44} = - x^{312} \,.
\]
Thus the Clifford algebra is built on basis vectors
\begin{eqnarray*}
       \varepsilon_0 &=& e_0\\
       \varepsilon_{i_1} &=& e_{i_1}\\
       \varepsilon_{i_2 i_1} &=&
       (e_{i_1}\otimes e_{i_2} - e_{i_2}\otimes e_{i_1})\\
       \hbox to 1cm{\dotfill}&&\hbox to 4cm{\dotfill} \\
       \varepsilon_{i_p \ldots i_2 i_1} &=&
       \sum\limits_\sigma \sign\sigma \cdot
       \sigma(e_{i_1}\otimes e_{i_2}\otimes\ldots\otimes e_{i_p}) \,,
\end{eqnarray*}
non-containing indices with identical values. These basis vectors obey the
multiplication rule
\begin{eqnarray}
       &&\varepsilon_{i_n...k_m...k_1...i_{p+1}} \circ
       \varepsilon_{i_p...k_m...k_1...i_1} \NN\\
       &&\qquad{}=
       \varepsilon_{i_n...i_1} \Cdot
       \prod\limits_{l=1}^m \sign\sigma_{i_p k_l} \Cdot
       \sign\sigma_{k_l i_{p+1}} \Cdot \g_{k_l k_l} \,,
\label{F05}
\end{eqnarray}
where $i$ enumerates distinct indices and $k$ enumerates conterminous indices
of comultipliers; $\sigma_{i_p k_l}$ is the permutation of index $i_p$ with
index $k_l$ in the second comultiplier, $\sigma_{k_l i_{p+1}}$ is the
permutation of index $k_l$ with index $i_{p+1}$ in the first
comultiplier.  For basis vectors with distinct indices
\[
       \varepsilon_{i_n\ldots i_{p+1}}\circ \varepsilon_{i_p \ldots i_2 i_1} =
       \varepsilon_{i_n \ldots i_2 i_1} \,.
\]

The space with basis vectors $\varepsilon_{i_p...i_1}$ will be denoted by
${\Bbb C}^p$.
If the dimensionality of the initial space $X$ is denoted by $n$, the
dimensionality of ${\Bbb C}^p$ is equal to number of
combinations of $n$ things $p$ at a time
\[
      \text{C}^p_n = \frac{n\Cdot(n-1)\Cdot\ldots\Cdot(n-p+1)}{p!} \,.
\]
Therefore
\[
      \text{dim}\,{\Bbb C}^p=\text{dim}\,{\Bbb C}^{n-p} \,, \quad
      \text{dim}\,{\Bbb C}^n=1 \,, \quad \text{dim}\,{\Bbb C}^{n-1}=n \,.
\]

The {\it space} of the Clifford algebra $\B{C}$ is a sum of spaces:
\[
       \B{C}={\Bbb C}^0+{\Bbb C}^1+\ldots+{\Bbb C}^n \,,
\]
where ${\Bbb C}^0=\bbbk$, ${\Bbb C}^1=X$.
The dimensionality of the Clifford algebra
\[
      N=\text{C}^0_n+\text{C}^1_n+\ldots+\text{C}^n_n=(1+1)^n=2^n.
\]
From here on the space $\B{C}$ will be identified with a {\it generalized
space-time of leptons}.

\subsection{Regular representation of Clifford algebra}

We consider further parastrophic matrices that represent the basis vectors
of the Clifford algebra $\B{C}$.

\subsubsection{Product of Clifford algebras}

First let us discuss the representation of product of Clifford algebras.

Hereinafter, when it is necessary to stress the dimensionality of Clifford
algebra we shall use the notation $\B{C}_n$, where $n$ is the dimensionality
of $X$.

The Clifford algebra $\B{C}_n$ can be written as the product
$\B{C}_m\times \B{C}_{n-m}$.  The representation of basis vectors of
$\B{C}_n$ in the subalgebra $\B{C}_m$ over the field of hypernumbers forming
algebra $\B{C}_{n-m}$ corresponds to this product.  Consider such a
representation for vector $x=\varepsilon_K\cdot x^K$.  The basis vectors
$\varepsilon_K$ can be written as
\[
     \varepsilon_K = \varepsilon_{k_2}\circ \varepsilon_{k_1}
     = \varepsilon_{d_1}\Cdot {\CC^{d_1}\!}_{k_1k_2} \,,
\]
where $\varepsilon_{k_1}$ are
basis vectors of subalgebra $\B{C}_m$, $\varepsilon_{k_2}$ are basis
vectors of subalgebra $\B{C}_{n-m}$, and
${\CC^{d_1}\!}_{k_1k_2}$ are the parastrophic matrices
of $\B{C}_{n-m}$ in $\B{C}_m$ over field of hypernumbers $\B{C}_{n-m}$.
We assume here that these parastrophic matrices can be expressed through
basis hypernumbers, $\xi_{k_2}$, of field $\B{C}_{n-m}$ as
\[
     {\CC^{d_1}\!}_{k_1k_2} = {\delta^{d_1}\!}_{k_1}\Cdot \xi_{k_2} \,.
\]
Then
\begin{equation}
     \varepsilon_{k_2}\circ \varepsilon_{k_1} =
     \varepsilon_{k_1}\Cdot \xi_{k_2} \,.
\label{F06}
\end{equation}
Thus the representation of vectors of algebra $\B{C}_n$ in the
subalgebra $\B{C}_m$ over field of hypernumbers $\B{C}_{n-m}$ has the form
\[
     x = \varepsilon_K\cdot x^K =
     \varepsilon_{k_2}\circ \varepsilon_{k_1}\Cdot x^{k_2k_1} =
     \varepsilon_{k_1}\Cdot (\xi_{k_2}\Cdot x^{k_2k_1}) \,.
\]
This representation will be called {\it complex} or {\it quaternion}, when
hypernumbers $\varepsilon_{k}$ are complex numbers or quaternions,
respectively.  The complex and quaternion representations considered below
are convenient by compactness.

\subsubsection{Classification of Clifford algebras}

We shall now elaborate a classification of Clifford algebras by means of
signatures of basis vectors.

Let us assign the basis vectors $\varepsilon_0$ and $\varepsilon_i$ to {\it
forming}, and the remaining basis vectors to {\it produced} bearing in mind
that these latter vectors are formed from $\varepsilon_0$ and $\varepsilon_i$
by $\circ$-multiplication.

The square of produced vector is expressed through the squares of forming
vectors. For example,
\[
\begin{array}{r@{}c@{}l@{}c@{}l}
       \varepsilon_{i_2 i_1 }         &{}\circ{}&
       \varepsilon_{i_2 i_1 }         &{}={}&
     {}- (\varepsilon_{i_1})^2 \Cdot
         (\varepsilon_{i_2})^2       \\
       \varepsilon_{i_3 i_2 i_1 }     &{}\circ{}&
       \varepsilon_{i_3 i_2 i_1 }     &{}={}&
     {}- (\varepsilon_{i_1})^2 \Cdot
         (\varepsilon_{i_2})^2 \Cdot
         (\varepsilon_{i_3})^2       \\
       \varepsilon_{i_4 i_3 i_2 i_1} &{}\circ{}&
       \varepsilon_{i_4 i_3 i_2 i_1} &{}={}&
      {}+ (\varepsilon_{i_1})^2 \Cdot
          (\varepsilon_{i_2})^2 \Cdot
          (\varepsilon_{i_3})^2 \Cdot
          (\varepsilon_{i_4})^2    \,.  \\
\end{array}
\]
But as $(\varepsilon_i)^2$ is equal either to $+\varepsilon_0$ or to
$-\varepsilon_0$ the Clifford algebras may be classified by the
signature of forming vector squares. Consider such a classification for
several cases of the dimensionality $n$ of space $X$.

(a) $n=0$, $N=1$, $\varepsilon_A=\{\varepsilon_0\}$. The signature of
square of $\varepsilon_0$ is
\[
     (+)\,.
\]

(b) $n=1$, $N=2$, $\varepsilon_A=\{\varepsilon_0,\varepsilon_1\}$. The two
variants of the signatures of forming vector squares are possible:
\[
     (+,+)\,,(+,-)\,.
\]
The last case is the algebra of complex numbers.

(c) $n=2$, $N=4$,
$\varepsilon_A=\{\varepsilon_0,\varepsilon_1,\varepsilon_2,\varepsilon_{21}\}$.
The three alternative sets of the signatures are possible:
\[
\begin{array}{l}
        ({}+{},{}+{}+{},{}-{}) \\
        ({}+{},{}+{}-{},{}+{}) \\
        ({}+{},{}-{}-{},{}-{}) \,.\\
\end{array}
\]
In the last case the Clifford algebra is called {\it the quaternion algebra}.

(d) $n=3$, $N=8$,
\[
     \varepsilon_A=\{\varepsilon_0,\varepsilon_1,\varepsilon_2,\varepsilon_3,
     \varepsilon_{21},\varepsilon_{13},\varepsilon_{32},\varepsilon_{123}\}\,.
\]
The possible variants of the signatures are
\[
\begin{array}{l}
        ({}+{},{}+{}+{}+{},{}-{}-{}-{},{}-{}) \\
        ({}+{},{}+{}+{}-{},{}-{}+{}+{},{}+{}) \\
        ({}+{},{}+{}-{}-{},{}+{}+{}-{},{}-{}) \\
        ({}+{},{}-{}-{}-{},{}-{}-{}-{},{}+{}) \,.\\
\end{array}
\]
The first case corresponds to the Clifford algebra constructed on the
geometric space.

(e) $n=4$, $N=16$,
\[
\begin{array}{@{}l@{}}
\varepsilon_A =
\{\varepsilon_0, \varepsilon_1, \varepsilon_2, \varepsilon_3, \varepsilon_4,
\varepsilon_{21}, \varepsilon_{13}, \varepsilon_{32},
\varepsilon_{14}, \varepsilon_{24}, \varepsilon_{34}, \quad \\
\multicolumn{1}{r}{
\varepsilon_{123}, \varepsilon_{124}, \varepsilon_{134}, \varepsilon_{234},
\varepsilon_{1234} \} \,.}
\end{array}
\]
The possible variants of the signatures are
\[
\begin{array}{@{}l@{}}
        ({}+{},{}+{}+{}+{}+{},{}-{}-{}-{}-{}-{}-{},{}-{}-{}-{}-{},{}+{}) \\
        ({}+{},{}+{}+{}+{}-{},{}-{}-{}-{}+{}+{}+{},{}-{}+{}+{}+{},{}-{}) \\
        ({}+{},{}+{}+{}-{}-{},{}-{}+{}+{}+{}+{}-{},{}+{}+{}-{}-{},{}+{}) \\
        ({}+{},{}+{}-{}-{}-{},{}+{}+{}-{}+{}-{}-{},{}-{}-{}-{}+{},{}-{}) \\
        ({}+{},{}-{}-{}-{}-{},{}-{}-{}-{}-{}-{}-{},{}+{}+{}+{}+{},{}+{}) \,.
\end{array}
\]
The second case corresponds to the Clifford algebra constructed on the
space-time.

We now formulate the rule of calculation of parastrophic matrices
for the Clifford algebra. This rule follows from the
regular representation of basis vectors:
\[
     \varepsilon_I \sim {\CC^L\!}_{KI} \equiv \CC_I\!\!\left({}^L_K\right)\,.
\]
Here $I$ is the parastrophic matrix number, $K$ is the column number, $L$ is
the row number. In order to calculate element ${\CC^L\!}_{KI}$, we proceed as
follows:  we evaluate the product $\varepsilon_K\circ\varepsilon_I$ by using
the multiplication rule \eref{F05} (the basis vector $\varepsilon_I$ is the
{\it right} comultiplier); the index $L$ of the resulting basis vector
$\varepsilon_L$ will be the row number for the desired element
${\CC^L\!}_{KI}$, and the coefficient at the specified basis vector will
determine its value.

Let us consider different representations of basis vectors of the
Clifford algebra $\B{C}$ when the initial space $X$ is: 1) the
geometric space, 2) the space-time.

\subsubsection{Clifford algebra on geometric space}
\label{Representation_features}

The algebra $\B{C}_3$ with signature $({}+{},{}+{}+{}+{},{}-{}-{}-{},{}-{})$
corresponds to this case.
Let us choose the special order of indices:
\[
     (32, 13, 21, 0, 1, 2, 3, 123) \,.
\]
Such a choice is justified by that the parastrophic matrices of the conjugate
Clifford algebra are represented by Pauli and Dirac matrices at this order of
indices (see Section~\ref{Pauli_Dirac_matrices}).  Using the above rule one
can obtain parastrophic matrices ${\CC^L\!}_{KI}$ (see Appendix~\ref{A1}).

The vector $x\in \B{C}_3$
can be expressed as
\begin{eqnarray}
      &&x =
     \varepsilon_{13}\circ (\varepsilon_{21}\Cdot x^{32} +
                             \varepsilon_0\Cdot x^{13}) \NN\\
      &&\quad{}+
     \varepsilon_0\circ (\varepsilon_{21}\Cdot x^{21} +
                         \varepsilon_0\Cdot x^0)
      {}+
     \varepsilon_2\circ (\varepsilon_{21}\Cdot x^1 +
                         \varepsilon_0\Cdot x^2) \NN\\
      &&\quad{}+
     \varepsilon_{123}\circ (\varepsilon_{21}\Cdot x^3 +
                             \varepsilon_0\Cdot x^{123}) \,.
\label{F07}
\end{eqnarray}
This decomposition corresponds to the representation of algebra $\B{C}_3$ as
product $\B{C}_2\times \B{C}_1$ and is complex.  The basis vectors of
$\B{C}_2$ are
$\varepsilon_{13},\varepsilon_0,\varepsilon_2,\varepsilon_{123}$; the basis
vectors of $\B{C}_1$ are $\varepsilon_{21},\varepsilon_0$.  The direction 21,
defining algebra $\B{C}_1$, will be called {\it basic}.  We put
the basis vector $\varepsilon_{21}$  into correspondence with
the imaginary unit $i$
bearing in mind that $(\varepsilon_{21})^2=-1$. The complex representation of
basis vectors $\varepsilon_A$ corresponding to \eref{F07} is given by
$4\times 4$ matrices (see Appendix~\ref{A1}), where the basis units ${\it 1}$, $i$,
$a$, $b$ replace the following blocks:
\begin{center}
\tighten
   ${\it 1} ={}\!$
   {\footnotesize
   \tabcolsep=1mm
   \renewcommand{\arraystretch}{0.95}
   \begin{tabular}{|c|c|}
     \hline
     \p1&    \\
     \hline
        &\p1 \\
     \hline
   \end{tabular}
   }$\,, \quad$
   $i ={}\!$
   {\footnotesize
   \tabcolsep=1mm
   \renewcommand{\arraystretch}{0.95}
   \begin{tabular}{|c|c|}
     \hline
        &\p1 \\
     \hline
     \m1&    \\
     \hline
   \end{tabular}
   }$\,, \quad$
   $a ={}\!$
   {\footnotesize
   \tabcolsep=1mm
   \renewcommand{\arraystretch}{0.95}
   \begin{tabular}{|c|c|}
     \hline
     \   &\p1 \\
     \hline
     \p1 &    \\
     \hline
   \end{tabular}
   }$\,, \quad$
   $b ={}\!$
   {\footnotesize
   \tabcolsep=1mm
   \renewcommand{\arraystretch}{0.95}
   \begin{tabular}{|c|c|}
     \hline
     \m1 &   \\
     \hline
         &\p1\\
     \hline
   \end{tabular}
   }$\,.$
\end{center}
It is significant that in the complex representation the matrix
\[
     {\CC^{i_1}\!}_{k_121} = i\Cdot {\delta^{i_1}\!}_{k_1}
\]
corresponds to the basis vector $\varepsilon_{21}$. As it will be shown in a
forthcoming paper \cite{art2}, this basic vector is closely connected with an
consideration of interaction between the lepton and the electromagnetic
field.

Note that $\varepsilon_{21}$ is the selected direction in the above
representation.  However the directions $\varepsilon_{13}$ and
$\varepsilon_{32}$ are equivalent to the direction $\varepsilon_{21}$ from an
algebraic point of view and can also be taken as basic.  In order to
distinguish these cases from the previous one, we denote imaginary unit by
$j$ when $\varepsilon_{13}$ is taken as basic direction, and by $k$ when
$\varepsilon_{32}$ is taken as basic direction.  In these cases, the regular
representation matrices differ from the matrices presented in
Appendix~\ref{A1} by replacement of the imaginary unit $i$ either on $j$, or
on $k$.

On the other hand, the vector $x\in\B{C}_3$ can be expressed in the
form of quaternion representation
\begin{eqnarray}
     &&x =
     (
     \varepsilon_{32}\Cdot x^{32} + \varepsilon_{13}\Cdot x^{13} +
     \varepsilon_{21}\Cdot x^{21} + \varepsilon_0\Cdot x^0)\circ
     \varepsilon_0 \NN\\
     &&\quad{}+(
     \varepsilon_{32}\Cdot x^1 + \varepsilon_{13}\Cdot x^2 +
     \varepsilon_{21}\Cdot x^3 + \varepsilon_0\Cdot x^{123})\circ
      \varepsilon_{123}\,.
\label{F08}
\end{eqnarray}
It corresponds to the representation of algebra $\B{C}_3$ as
product $\B{C}_1\times\B{C}_2$. The basis vectors of $\B{C}_1$ are
$\varepsilon_0,\varepsilon_{123}$; the basis vectors of $\B{C}_2$ are
$\varepsilon_{32},\varepsilon_{13},\varepsilon_{21},\varepsilon_0$.
The quaternion representation of the basis vectors $\varepsilon_A$
corresponding to \eref{F08} is given by $2\times 2$ matrices presented also
in Appendix~\ref{A1}.

\subsubsection{Clifford algebra on space-time}

The Clifford algebra $\B{C}_4$ with the signature
\[
     ({}+{},{}+{}+{}+{}-{},{}-{}-{}-{}+{}+{}+{},{}-{}+{}+{}+{},{}-{})
\]
corresponds to this case.
Choose the special order of indices:
\[
     (32, 13, 21, 0, 42, 14, 1324, 34, 1, 2, 3, 123, 134, 234, 4, 124) \,.
\]
Now one can calculate the parastrophic matrices ${\CC^L\!}_{KI}$ by using
the rule formulated above.  They are presented in Appendix~\ref{A2}.

The complex representation of the basis vectors $\varepsilon_I$,
corresponding to the decomposition:
\begin{eqnarray*}
      &&x =
     \varepsilon_{13}\circ (\varepsilon_{21}\Cdot x^{32} +
                             \varepsilon_0\Cdot x^{13}) \\
      &&{}+
     \varepsilon_0\circ (\varepsilon_{21}\Cdot x^{21} +
                         \varepsilon_0\Cdot x^0) +
     \varepsilon_{14}\circ (\varepsilon_{21}\Cdot x^{42} +
                             \varepsilon_0\Cdot x^{14}) \\
      &&{}+
     \varepsilon_{34}\circ (\varepsilon_{21}\Cdot x^{1324} +
                         \varepsilon_0\Cdot x^{34}) +
     \varepsilon_2\circ (\varepsilon_{21}\Cdot x^1 +
                         \varepsilon_0\Cdot x^2) \\
      &&{}+
     \varepsilon_{123}\circ (\varepsilon_{21}\Cdot x^3 +
                             \varepsilon_0\Cdot x^{123})
      {}+
     \varepsilon_{234}\circ (\varepsilon_{21}\Cdot x^{134} +
                             \varepsilon_0\Cdot x^{234}) \\
      &&{}+
     \varepsilon_{124}\circ (\varepsilon_{21}\Cdot x^4 +
                         \varepsilon_0\Cdot x^{124}) \,,
\end{eqnarray*}
is given by $8\times 8$ matrices where the above blocks are replaced by the
basis units ${\it 1}$, $i$, $a$, $b$ (see Appendix~\ref{A2}).

The quaternion representation of the vectors $\varepsilon_I$, corresponding
to decomposition:
\begin{eqnarray*}
     &&x {}=
     (
     \varepsilon_{32}\Cdot x^{32} + \varepsilon_{13}\Cdot x^{13} +
     \varepsilon_{21}\Cdot x^{21} + \varepsilon_0\Cdot x^0)\circ
     \varepsilon_0\\
     &&\quad{}+
     (
     \varepsilon_{32}\Cdot x^{42} + \varepsilon_{13}\Cdot x^{14} +
     \varepsilon_{21}\Cdot x^{1324} + \varepsilon_0\Cdot x^{34})\circ
     \varepsilon_{34}\\
     &&\quad{}+
     (
     \varepsilon_{32}\Cdot x^1 + \varepsilon_{13}\Cdot x^2 +
     \varepsilon_{21}\Cdot x^3 + \varepsilon_0\Cdot x^{123})\circ
     \varepsilon_{123}\\
     &&\quad{}+
     (
     \varepsilon_{32}\Cdot x^{134} + \varepsilon_{13}\Cdot x^{234} +
     \varepsilon_{21}\Cdot x^4 + \varepsilon_0\Cdot x^{124})\circ
     \varepsilon_{124} \,,
\end{eqnarray*}
is given by $4\times 4$ matrices presented also in Appendix~\ref{A2}.

\subsection{Generalized conjugate space-time. Conjugate Clifford algebra.}

To describe the space structure of antiparticles we introduce a {\it
generalized conjugate space-time}.  For this purpose we generalize the
conjugate space-time $\XT$ introduced in Section~\ref{Conjugate_space} up to
{\it universal algebra of covariant tensors} $\BT{X}$ in line with
Section~\ref{Gen-Space-Time}.  The vector $\tld{x}\in \BT{X}$ can be
expressed through basis vectors
\begin{eqnarray*}
      \tld{x} =
      x_0\Cdot E^0  &&{}+
            x_{i_1}\Cdot E^{i_1}   +
            x_{i_2 i_1}\Cdot E^{i_1 i_2}  + \ldots \\
       &&{}+
      x_{i_n ...i_2i_1}\Cdot E^{i_1 i_2...i_n}  + \ldots = x_I\cdot E^I  \,.
\end{eqnarray*}
Here $E^0$ is the unit of reals $\bbbk$. Note that the
$\otm$-multiplication in $\BT{X}$ differs from multiplication in $\B{X}$ by
an order of multipliers.

The convolution of linear spaces $X$ and $\XT$ is generalized up to
convolution of linear spaces $\B{X}$ and $\BT{X}$. In particular the basis
vectors $e_{k_n...k_2k_1}$ and $E^{i_1i_2...i_n}$ can be chosen so that
\[
     \la E^{i_1i_2...i_n},e_{k_n...k_2k_1}\ra =
     \delta_{k_n}^{i_n}\Cdot\ldots\Cdot
     \delta_{k_2}^{i_2}\Cdot\delta_{k_1}^{i_1}\,.
\]

Let $\BT{D}$ be subalgebra of $\BT{X}$. Then
the rule of multiplication of the basis vectors is written as
\[
     E^I\circ E^K = {\CC^{IK}\!}_L\cdot E^L \,,
\]
where ${\CC^{IK}\!}_L$ are the structure constants of the conjugate subalgebra
$\BT{D}$. In a specific case, the multiplication $E^I\circ E^K$ defines
the convolution
\[
     \la E^I,E^K\ra = {\CC^{IK}\!}_0\Cdot E^0
\]
and the inverse metric tensor $\g^{IK}={\CC^{IK}\!}_0$. Note also that
${\CC^{0K}\!}_I={\CC^{K0}\!}_I={\delta^K\!}_I$.

From the condition of associativity of multiplication
\[
    E^I\circ(E^K\circ E^N) = (E^I\circ E^K)\circ E^N \,,
\]
it follows that
\[
     {\CC^{KN}\!}_L\cdot {\CC^{IL}\!}_M ={\CC^{IK}\!}_L\cdot {\CC^{LN}\!}_M\,.
\]
This expression demonstrates the possibility of regular representation of
basis vectors $E^I$:
\[
     E^I \sim {\CC^{IL}\!}_M \,.
\]

The relation between the structure constants of algebra $\BT{D}$ and of
algebra $\B{D}$ is given by the operation of {\it conjugation}:
\begin{equation}
     {\CC^{RQ}\!}_P = \g^{RI} \cdot \g^{QK} \cdot {\CC^L\!}_{KI} \cdot \g_{LP} \,.
\label{F10}
\end{equation}

A {\it conjugate Clifford algebra} $\BT{C}$ can be define as subalgebra of
universal covariant algebra $\BT{X}$ by analogy with
the Clifford algebra $\B{C}$. From here on the space $\BT{C}$ will be
identified with a {\it generalized space-time of antileptons}. The basis
vectors of $\BT{C}$ will be denoted by $\Epsilon^{i_1 i_2\ldots i_n}$.
In particular, for the basis vectors with distinct indices
\[
       \Epsilon^{i_1 i_2\ldots i_p}\circ \Epsilon^{i_{p+1} \ldots i_n} =
       \Epsilon^{i_1 i_2\ldots i_n} \,.
\]

\subsection{Regular representation of conjugate Clifford
algebra. Pauli and Dirac matrices}
\label{Pauli_Dirac_matrices}

From the regular representation of basis vectors
\[
     \Epsilon^I \sim {\CC^{IK}\!}_L \equiv \CC^I\!\!\left({}^K_L\right) \,,
\]
the calculation rule of parastrophic matrices for the conjugate Clifford
algebra follows.  Here $I$ is the parastrophic matrix number, $L$ is the
column number, $K$ is the row number. In order to calculate element
${\CC^{IK}\!}_L$, we proceed as follows: we evaluate the product
$\Epsilon^I\circ\Epsilon^K$ by using the multiplication rule of the type
\eref{F05} (the basis vector $\Epsilon^I$ is the {\it left} comultiplier);
the index $L$ of the resulting basis vector $\Epsilon_L$ will be the column
number for the desired element ${\CC^{IK}\!}_L$, and the coefficient at the
specified basis vector will determine its value. The resulting matrix must be
multiplied by $(\Epsilon^I)^2$.

Consider different representations of basis vectors of the conjugate
Clifford algebra $\BT{C}$ when the initial space $\XT$ is: 1) the
geometric space ($\XT_3\equiv X_3$), 2) the conjugate space-time.

\subsubsection{Conjugate Clifford algebra on geometric space}

The algebra $\BT{C}_3$ with the signature
$({}+{},{}+{}+{}+{},{}-{}-{}-{},{}-{})$ corresponds to this case.  Choose the
index order applied previously: $(32, 13, 21, 0, 1, 2, 3, 123)$. The
parastrophic matrices ${\CC_L\!}^{KI}$ can be calculated as from
the above rule so and from the relation \eref{F10}.  In conjugate Clifford
algebra $\BT{C}_3$ the metric tensor is
\begin{center}
\tighten
   \raisebox{-2.3mm}{$\g^{LP}\,\sim\;$}
   {\footnotesize
   \tabcolsep=1mm
   \renewcommand{\arraystretch}{0.95}
   \begin{tabular}{c|cccc|cccc|}
         \multicolumn{2}{r}{}&
         \ii{13}&&
         \multicolumn{1}{c}{\ii{ 0}}&&
         \ii{ 2}&&
         \multicolumn{1}{c}{\ii{123}}\\
         \vsp{-2mm}
         \multicolumn{1}{l}{}&
         \ii{32}&&\ii{21}&
         \multicolumn{1}{l}{}&
         \ii{ 1}&&\ii{ 3}\\
   \cline{2-9}
   \ib{32} &\m1&   &   &   &   &   &   &    \\
   \ib{13} &   &\m1&   &   &   &   &   &    \\
   \ib{21} &   &   &\m1&   &   &   &   &    \\
   \ib{0}  &   &   &   &\p1&   &   &   &    \\
   \cline{2-9}
   \ib{1}  &   &   &   &   &\p1&   &   &    \\
   \ib{2}  &   &   &   &   &   &\p1&   &    \\
   \ib{3}  &   &   &   &   &   &   &\p1&    \\
   \ib{123}&   &   &   &   &   &   &   &\m1 \\
   \cline{2-9}
   \end{tabular}}%
\end{center}
If we substitute this tensor and the parastrophic matrices
of Clifford algebra $\B{C}_3$ (see Appendix~\ref{A1}) in \eref{F10},
we find the regular representation matrices of basis vectors
for conjugate Clifford algebra $\BT{C}_3$.

The eight parastrophic matrices ${\CC_L\!}^{KI}$ of algebra $\BT{C}_3$
are written in the real, complex, and quaternion representations in
Appendix~\ref{A3}. The complex representation is based on the
decomposition
\begin{eqnarray*}
     &&\tld{x} =
     \Epsilon^{13}\circ  (x_{32}\Cdot \Epsilon^{21} +
                         x_{13}\Cdot \Epsilon^0)+
     \Epsilon^0\circ (x_{21}\Cdot \Epsilon^{21} +
                         x_0\Cdot \Epsilon^0) \\
     &&{}+
     \Epsilon^2\circ (x_1\Cdot \Epsilon^{21} +
                         x_2\Cdot \Epsilon^0)+
     \Epsilon^{123}\circ (x_3\Cdot \Epsilon^{21} +
                         x_{123}\Cdot \Epsilon^0) \,.
\end{eqnarray*}
The quaternion representation is based on the decomposition
\begin{eqnarray*}
     &&\tld{x} =
     (
     x_{32}\Cdot \Epsilon^{32} + x_{13}\Cdot \Epsilon^{13} +
     x_{21}\Cdot \Epsilon^{21} + x_0\Cdot \Epsilon^0)\circ\Epsilon^0 \\
     &&{}+(x_1\Cdot \Epsilon^{32} + x_2\Cdot \Epsilon^{13} +
     x_3\Cdot \Epsilon^{21} + x_{123}\Cdot \Epsilon^0)\circ\Epsilon^{123} \,.
\end{eqnarray*}
In quaternion representation the parastrophic matrices contain the Pauli
matrices $\sigma_1$, $\sigma_2$, $\sigma_3$ as basis blocks.

Let us introduce matrices $\gamma_0$, $\gamma_1$, $\gamma_2$, $\gamma_3$,
$\gamma_{21}$, $\gamma_{13}$, $\gamma_{32}$ and $\gamma_{123}$ through
the relations
\begin{eqnarray*}
    &&\Epsilon^0 = {\it 1}\Cdot \gamma_0 \,,\qquad
    \Epsilon^1 = i\Cdot \gamma_1 \,,\qquad
    \Epsilon^2 = i\Cdot \gamma_2 \,,\qquad
    \Epsilon^3 = i\Cdot \gamma_3 \,,\\
    &&\Epsilon^{21} = \gamma_1\Cdot\gamma_2=\gamma_{12} \,,\qquad
    \Epsilon^{13} = \gamma_3\Cdot\gamma_1=\gamma_{31} \,,\\
    &&\Epsilon^{32} = \gamma_2\Cdot\gamma_3=\gamma_{23} \,,\qquad
    \Epsilon^{123} = (-i) \Cdot \gamma_1 \Cdot \gamma_2 \Cdot \gamma_3 =
    (-i) \Cdot \gamma_{123}\,.
\end{eqnarray*}
These matrices
\[
\begin{tabular}{lll}
\tighten
   $\gamma_0 ={}\!\!$
   {\footnotesize
   \tabcolsep=1mm
   \renewcommand{\arraystretch}{0.95}
   \begin{tabular}{|c|c|}
     \hline
     \Pl{\openone}&              \\
     \hline
                  &\Pl{\openone} \\
     \hline
   \end{tabular}
   }$\,, \quad$ &
\tighten
   $\gamma_{123}= i \Cdot {}\!\!\!$
   {\footnotesize
   \tabcolsep=1mm
   \renewcommand{\arraystretch}{0.95}
   \begin{tabular}{|c|c|}
     \hline
                  &\ml{\openone} \\
     \hline
     \Pl{\openone}&              \\
     \hline
   \end{tabular}
   }$\,,$\\[5mm]
\tighten
   $\gamma_1={}\!\!$
   {\footnotesize
   \tabcolsep=1mm
   \renewcommand{\arraystretch}{0.95}
   \begin{tabular}{|c|c|}
     \hline
                  &\mL{\sigma_1} \\
     \hline
     \pL{\sigma_1}&              \\
     \hline
   \end{tabular}
   }$\,, \quad$ &
\tighten
   $\gamma_2={}\!\!$
   {\footnotesize
   \tabcolsep=1mm
   \renewcommand{\arraystretch}{0.95}
   \begin{tabular}{|c|c|}
     \hline
                  &\mL{\sigma_2} \\
     \hline
     \pL{\sigma_2}&              \\
     \hline
   \end{tabular}
   }$\,, \quad$ &
\tighten
   $\gamma_3={}\!\!$
   {\footnotesize
   \tabcolsep=1mm
   \renewcommand{\arraystretch}{0.95}
   \begin{tabular}{|c|c|}
     \hline
                  &\mL{\sigma_3} \\
     \hline
     \pL{\sigma_3}&              \\
     \hline
   \end{tabular}
   }$\,,$\\[5mm]
\tighten
   $\gamma_{12}= i \Cdot {}\!\!\!$
   {\footnotesize
   \tabcolsep=1mm
   \renewcommand{\arraystretch}{0.95}
   \begin{tabular}{|c|c|}
     \hline
     \Pl{\sigma_3}&              \\
     \hline
                  &\Pl{\sigma_3} \\
     \hline
   \end{tabular}
   }$\,, \quad$ &
\tighten
   $\gamma_{31}= i \Cdot {}\!\!\!$
   {\footnotesize
   \tabcolsep=1mm
   \renewcommand{\arraystretch}{0.95}
   \begin{tabular}{|c|c|}
     \hline
     \Pl{\sigma_2}&              \\
     \hline
                  &\Pl{\sigma_2} \\
     \hline
   \end{tabular}
   }$\,, \quad$ &
\tighten
   $\gamma_{23}= i \Cdot {}\!\!\!$
   {\footnotesize
   \tabcolsep=1mm
   \renewcommand{\arraystretch}{0.95}
   \begin{tabular}{|c|c|}
     \hline
     \Pl{\sigma_1}&              \\
     \hline
                  &\Pl{\sigma_1} \\
     \hline
   \end{tabular}
   }
\end{tabular}
\]
form the incomplete set of {\it Dirac matrices} describing only
spatial basis vectors.

\subsubsection{Conjugate Clifford algebra on conjugate space-time}

The Clifford algebra $\BT{C}_4$ with signature
\[
     ({}+{},{}+{}+{}+{}-{},{}-{}-{}-{}+{}+{}+{},{}-{}+{}+{}+{},{}-{})
\]
corresponds to this case.
The parastrophic matrices ${\CC^{IK}\!}_L$ are calculated for
the index order applied previously
\[
     (32, 13, 21, 0, 42, 14, 1324, 34, 1, 2, 3, 123, 134, 234, 4, 124) \,.
\]
The regular representation of basis
vectors for conjugate Clifford algebra $\BT{C}_4$
can also be obtained from relation \eref{F10} by using
the metric tensor
\begin{center}
\tighten
   \raisebox{-2.5mm}{$\g^{LP} \,\sim\;$}
   {\footnotesize
   \tabcolsep=1mm
   \renewcommand{\arraystretch}{0.95}
   \begin{tabular}{c|cccc|cccc|cccc|cccc|}
         \multicolumn{2}{r}{}&
         \ii{13}&&
         \multicolumn{1}{c}{\ii{ 0}}&&
         \ii{14}&&
         \multicolumn{1}{c}{\ii{34}}&&
         \ii{ 2}&&
         \multicolumn{1}{c}{\ii{123}}&&
         \ii{234}&&
         \multicolumn{1}{c}{\ii{124}}\\
         \vsp{-1.5mm}
         \multicolumn{1}{l}{}&
         \ii{32}&&\ii{21}&
         \multicolumn{1}{c}{}&
         \ii{42}&&\ii{1324}&
         \multicolumn{1}{c}{}&
         \ii{1}&&\ii{3}&
         \multicolumn{1}{c}{}&
         \ii{134}&&\ii{4}  \\
   \cline{2-17}
     \ib{32}  &\m1&   &   &   &   &   &   &   &   &   &   &   &   &   &   &    \\
     \ib{13}  &   &\m1&   &   &   &   &   &   &   &   &   &   &   &   &   &    \\
     \ib{21}  &   &   &\m1&   &   &   &   &   &   &   &   &   &   &   &   &    \\
     \ib{0}   &   &   &   &\p1&   &   &   &   &   &   &   &   &   &   &   &    \\
   \cline{2-17}
     \ib{42}  &   &   &   &   &\p1&   &   &   &   &   &   &   &   &   &   &    \\
     \ib{14}  &   &   &   &   &   &\p1&   &   &   &   &   &   &   &   &   &    \\
     \ib{1324}&   &   &   &   &   &   &\m1&   &   &   &   &   &   &   &   &    \\
     \ib{34}  &   &   &   &   &   &   &   &\p1&   &   &   &   &   &   &   &    \\
   \cline{2-17}
     \ib{1}   &   &   &   &   &   &   &   &   &\p1&   &   &   &   &   &   &    \\
     \ib{2}   &   &   &   &   &   &   &   &   &   &\p1&   &   &   &   &   &    \\
     \ib{3}   &   &   &   &   &   &   &   &   &   &   &\p1&   &   &   &   &    \\
     \ib{123} &   &   &   &   &   &   &   &   &   &   &   &\m1&   &   &   &    \\
   \cline{2-17}
     \ib{134} &   &   &   &   &   &   &   &   &   &   &   &   &\p1&   &   &    \\
     \ib{234} &   &   &   &   &   &   &   &   &   &   &   &   &   &\p1&   &    \\
     \ib{4}   &   &   &   &   &   &   &   &   &   &   &   &   &   &   &\m1&    \\
     \ib{124} &   &   &   &   &   &   &   &   &   &   &   &   &   &   &   &\p1 \\
   \cline{2-17}
   \end{tabular}}%
\end{center}

16 parastrophic matrices of algebra $\BT{C}_4$ are written in the real,
complex, and quaternion representations in Appendix~\ref{A3}.  The
complex representation of the basis vectors is based on decomposition
of vector $\tld{x}\in\BT{C}_4$:
\begin{eqnarray*}
     &&\tld{x} =
     \Epsilon^{13}\circ (x_{32}\Cdot \Epsilon^{21} +
                         x_{13}\Cdot \Epsilon^0) \\
     &&{}+
     \Epsilon^0\circ (x_{21}\Cdot \Epsilon^{21} +
                         x_0\Cdot \Epsilon^0)
     {}+
     \Epsilon^{14}\circ (x_{42}\Cdot \Epsilon^{21} +
                         x_{14}\Cdot \Epsilon^0) \\
     &&{}+
     \Epsilon^{34}\circ (x_{1324}\Cdot \Epsilon^{21} +
                         x_{34}\Cdot \Epsilon^0)
     {}+
     \Epsilon^2\circ (x_1\Cdot \Epsilon^{21} +
                      x_2\Cdot \Epsilon^0) \\
     &&{}+
     \Epsilon^{123}\circ (x_3\Cdot \Epsilon^{21} +
                          x_{123}\Cdot \Epsilon^0)
     {}+
     \Epsilon^{234}\circ (x_{134}\Cdot \Epsilon^{21} +
                          x_{234}\Cdot \Epsilon^0) \\
     &&{}+
     \Epsilon^{124}\circ (x_4\Cdot \Epsilon^{21} +
                          x_{124}\Cdot \Epsilon^0)\,.
\end{eqnarray*}
It is given by $8\times 8$ matrices where blocks are replaced by the basis
units ${\it 1}$ and $i$.

The quaternion representation of vectors $\Epsilon^I$ corresponds to
decomposition:
\begin{eqnarray*}
     &&\tld{x} =
     (
     x_{32}\Cdot \Epsilon^{32} + x_{13}\Cdot \Epsilon^{13} +
     x_{21}\Cdot \Epsilon^{21} + x_0\Cdot \Epsilon^0)\circ\Epsilon^0 \\
     &&{}+
     (
     x_{42}\Cdot \Epsilon^{32} + x_{14}\Cdot \Epsilon^{13} +
     x_{1324}\Cdot \Epsilon^{21} + x_{34}\Cdot \Epsilon^0)\circ\Epsilon^{34} \\
     &&{}+
     (
     x_1\Cdot \Epsilon^{32} + x_2\Cdot \Epsilon^{13} +
     x_3\Cdot \Epsilon^{21} + x_{123}\Cdot \Epsilon^0)\circ\Epsilon^{123} \\
     &&{}+
     (
     x_{134}\Cdot \Epsilon^{32} + x_{234}\Cdot \Epsilon^{13} +
     x_4\Cdot \Epsilon^{21} + x_{124}\Cdot \Epsilon^0)\circ\Epsilon^{124} \,.
\end{eqnarray*}
It is given by $4\times 4$ matrices.

\subsection{Approximate representation of basis vectors}
\label{Approx_Representation}

In this Section consider the regular representation of basis
vectors of algebra $\B{C}_n$ in its subalgebra $\B{C}_{n-k}$ for $k<n$.
Such a representation will be called {\it approximate}.
Further it will be used to obtain the Dirac and Pauli matrices.

For example, consider the approximate representation of basis vectors of
algebra $\BT{C}_n$ in its subalgebra $\BT{C}_{n-1}$.

Let us separate the basis vectors $\Epsilon^I$ of algebra $\BT{C}_n$ into two
groups $\Epsilon^{I_1}$ and $\Epsilon^{I_2}$ with the same number of vectors
so that the vectors $\Epsilon^{I_1}$ make up algebra.  Because of
symmetries of the Clifford algebra, the relations \eref{F03} take the form
\begin{eqnarray}
       \Epsilon^{I_1}\circ \Epsilon^{K_1} &=&
       {\CC^{I_1K_1}\!}_{L_1}\cdot \Epsilon^{L_1} \,,
\label{F11}
     \\
       \Epsilon^{I_2}\circ \Epsilon^{K_1} &=&
       {\CC^{I_2K_1}\!}_{L_2}\cdot \Epsilon^{L_2} \,,
\label{F12}
     \\
       \Epsilon^{I_1}\circ \Epsilon^{K_2} &=&
       {\CC^{I_1K_2}\!}_{L_2}\cdot \Epsilon^{L_2} \,,\NN\\
       \Epsilon^{I_2}\circ \Epsilon^{K_2} &=&
       {\CC^{I_2K_2}\!}_{L_1}\cdot \Epsilon^{L_1} \,.\NN
\end{eqnarray}
We assume that, approximately by calculating the representation
matrices of basis vectors of $\BT{C}_n$ in $\BT{C}_{n-1}$, the basis vectors
$\Epsilon^{L_2}$ can be replaced by $\Epsilon^{L_1}$ through the relation:
\[
     \Epsilon^{L_2} = {P^{L_2}\!}_{L_1} \cdot \Epsilon^{L_1} \,,
\]
where ${P^{L_2}\!}_{L_1}$ is the correspondence matrix.
Then the relation \eref{F12} takes the form
\begin{equation}
     \Epsilon^{I_2}\circ \Epsilon^{K_1} =
     {\CC^{I_2K_1}\!}_{L_2}\cdot {P^{L_2}\!}_{L_1} \cdot \Epsilon^{L_1} \,.
\label{F13}
\end{equation}
From \eref{F11} and \eref{F13} the representation matrices of basis
vectors of the algebra $\BT{C}_n$ in its subalgebra $\BT{C}_{n-1}$ can be
calculated. Note that the basis vectors of $\BT{C}_{n-1}$ are exactly
represented but the other basis vectors are approximately represented.

Similarly it is possible to consider the approximate representation of basis
vectors of algebra $\B{C}_n$ in its subalgebra $\B{C}_{n-k}$, where $k<n$.

\subsubsection{First approximate representation}

Consider the approximate representation of basis vectors of algebra
$\BT{C}_4$ in its subalgebra $\BT{C}_3$ constructed on the basis vectors
$\Epsilon^{32}$, $\Epsilon^{13}$, $\Epsilon^{21}$, $\Epsilon^0$,
$\Epsilon^1$, $\Epsilon^2$, $\Epsilon^3$, $\Epsilon^{123}$.
For this purpose, we assume that by calculating the parastrophic matrices
by \eref{F13} the basis vectors with indices
\[
     42,14, 1324,34, 134,234, 4,124
\]
are replaced by the basis vectors with indices
\[
     32,13, 21,0, 1,2,3,123
\]
respectively.
Then the dimensionality of matrices of algebra $\BT{C}_4$ is reduced by half
and equal to $8\times 8$ for the real representation, $4\times 4$ for the
complex representation, and $2\times 2$ for the quaternion representation.
For example,
{\renewcommand{\arraystretch}{0.95}
\tighten
\begin{flushleft}
\tabcolsep=1.5mm
\hspace*{0mm}%
\begin{tabular}{@{}lrr@{}}
   \raisebox{-4.9mm}{$\Epsilon^{4} \,\sim (-1)\Cdot{}\!\!\!$}&
   {\footnotesize
   \tabcolsep=1mm
   \begin{tabular}{c|cccc|cccc|}
         \multicolumn{2}{r}{}&
         \ii{\it 14}&&
         \multicolumn{1}{c}{\ii{\it 34}}&&
         \ii{\it 234}&&
         \multicolumn{1}{c}{\ii{\it 124}}\\
         \vsp{-1mm}
         \multicolumn{1}{l}{}&
         \ii{\it 42}&&\ii{\it 1324}&
         \multicolumn{1}{l}{}&
         \ii{\it 134}&&\ii{\it  4}\\
         \vsp{0mm}
         \multicolumn{2}{r}{}&
         \ii{13}&&
         \multicolumn{1}{c}{\ii{ 0}}&&
         \ii{ 2}&&
         \multicolumn{1}{c}{\ii{123}}\\
         \vsp{-2mm}
         \multicolumn{1}{l}{}&
         \ii{32}&&\ii{21}&
         \multicolumn{1}{l}{}&
         \ii{ 1}&&\ii{ 3}\\
   \cline{2-9}
   \ib{32}  &   &   &   &   &   &\m1&   &    \\
   \ib{13}  &   &   &   &   &\p1&   &   &    \\
   \ib{21}  &   &   &   &   &   &   &   &\m1 \\
   \ib{0}   &   &   &   &   &   &   &\p1&    \\
   \cline{2-9}
   \ib{1}   &   &\m1&   &   &   &   &   &    \\
   \ib{2}   &\p1&   &   &   &   &   &   &    \\
   \ib{3}   &   &   &   &\m1&   &   &   &    \\
   \ib{123} &   &   &\p1&   &   &   &   &    \\
   \cline{2-9}
   \end{tabular}
   }%
&
\begin{tabular}{r}
   \vsp{10.3mm}
   =\hspace*{2\tabcolsep}%
   {$i \,\Cdot$
   \footnotesize
   \tabcolsep=1mm
   \begin{tabular}{|cc|cc|}
     \hline
        &   &\p1&    \\
        &   &   &\p1 \\
     \hline
     \p1&   &   &    \\
        &\p1&   &    \\
     \hline
   \end{tabular}
   }%
   \\
   \vsp{4.1mm}%
   =\hspace*{2\tabcolsep}%
   {$i \,\Cdot$
   \tabcolsep=1mm
   \begin{tabular}{|c|c|}
     \hline
                  &\Pl{\openone} \\
     \hline
     \Pl{\openone}&              \\
     \hline
   \end{tabular}
   }%
\end{tabular}
\end{tabular}
\end{flushleft}
}%
\noindent
As a result, we obtain
\begin{center}
\tighten
   $\Epsilon^{0} \sim{}$
   ${\it 1}\Cdot{}\!\!\!$
   {\footnotesize
   \tabcolsep=1mm
   \renewcommand{\arraystretch}{0.95}
   \begin{tabular}{|c|c|}
     \hline
     \Pl{\openone}&              \\
     \hline
                  &\Pl{\openone} \\
     \hline
   \end{tabular}
   }$\,, \quad$
   $\Epsilon^{1324} \sim{}$
   $i \Cdot{}\!\!\!$
   {\footnotesize
   \tabcolsep=1mm
   \renewcommand{\arraystretch}{0.95}
   \begin{tabular}{|c|c|}
     \hline
     \Pl{\openone}&              \\
     \hline
                  &\ml{\openone} \\
     \hline
   \end{tabular}
   }$\,, \quad$
\end{center}
\begin{center}
\tighten
   $\Epsilon^{1} \sim{}$
   $i \Cdot{}\!\!\!$
   {\footnotesize
   \tabcolsep=1mm
   \renewcommand{\arraystretch}{0.95}
   \begin{tabular}{|c|c|}
     \hline
                  &\mL{\sigma_1} \\
     \hline
     \pL{\sigma_1}&              \\
     \hline
   \end{tabular}
   }$\,, \quad$
   $\Epsilon^{2} \sim{}$
   $i \Cdot{}\!\!\!$
   {\footnotesize
   \tabcolsep=1mm
   \renewcommand{\arraystretch}{0.95}
   \begin{tabular}{|c|c|}
     \hline
                  &\mL{\sigma_2} \\
     \hline
     \pL{\sigma_2}&              \\
     \hline
   \end{tabular}
   }$\,, \quad$   \\
\end{center}
\begin{center}
\tighten
   $\Epsilon^{3} \sim{}$
   $i \Cdot{}\!\!\!$
   {\footnotesize
   \tabcolsep=1mm
   \renewcommand{\arraystretch}{0.95}
   \begin{tabular}{|c|c|}
     \hline
                  &\mL{\sigma_3} \\
     \hline
     \pL{\sigma_3}&              \\
     \hline
   \end{tabular}
   }$\,, \quad$
   $\Epsilon^{4} \sim{}$
   $ i \Cdot{}\!\!\!$
   {\footnotesize
   \tabcolsep=1mm
   \renewcommand{\arraystretch}{0.95}
   \begin{tabular}{|c|c|}
     \hline
                  &\Pl{\openone} \\
     \hline
     \Pl{\openone}&              \\
     \hline
   \end{tabular}
   }$\,, \quad$         \\
\end{center}
\begin{center}
\tighten
   $\Epsilon^{21} \sim{}$
   $ i \Cdot{}\!\!\!$
   {\footnotesize
   \tabcolsep=1mm
   \renewcommand{\arraystretch}{0.95}
   \begin{tabular}{|c|c|}
     \hline
     \Pl{\sigma_3}&              \\
     \hline
                  &\Pl{\sigma_3} \\
     \hline
   \end{tabular}
   }$\,, \quad$
   $\Epsilon^{13} \sim{}$
   $ i \Cdot{}\!\!\!$
   {\footnotesize
   \tabcolsep=1mm
   \renewcommand{\arraystretch}{0.95}
   \begin{tabular}{|c|c|}
     \hline
     \Pl{\sigma_2}&              \\
     \hline
                  &\Pl{\sigma_2} \\
     \hline
   \end{tabular}
   }$\,, \quad$   \\
\end{center}
\begin{center}
\tighten
   $\Epsilon^{32} \sim{}$
   $ i \Cdot{}\!\!\!$
   {\footnotesize
   \tabcolsep=1mm
   \renewcommand{\arraystretch}{0.95}
   \begin{tabular}{|c|c|}
     \hline
     \Pl{\sigma_1}&              \\
     \hline
                  &\Pl{\sigma_1} \\
     \hline
   \end{tabular}
   }$\,, \quad$
   $\Epsilon^{14} \sim{}$
   ${\it 1}\Cdot{}\!\!\!$
   {\footnotesize
   \tabcolsep=1mm
   \renewcommand{\arraystretch}{0.95}
   \begin{tabular}{|c|c|}
     \hline
     \pL{\sigma_1}&              \\
     \hline
                  &\mL{\sigma_1} \\
     \hline
   \end{tabular}
   }$\,, \quad$   \\
\end{center}
\begin{center}
\tighten
   $\Epsilon^{42} \sim{}$
   ${\it 1}\Cdot{}\!\!\!$
   {\footnotesize
   \tabcolsep=1mm
   \renewcommand{\arraystretch}{0.95}
   \begin{tabular}{|c|c|}
     \hline
     \mL{\sigma_2}&              \\
     \hline
                  &\pL{\sigma_2} \\
     \hline
   \end{tabular}
   }$\,, \quad$
   $\Epsilon^{34} \sim{}$
   ${\it 1}\Cdot{}\!\!\!$
   {\footnotesize
   \tabcolsep=1mm
   \renewcommand{\arraystretch}{0.95}
   \begin{tabular}{|c|c|}
     \hline
     \pL{\sigma_3}&              \\
     \hline
                  &\mL{\sigma_3} \\
     \hline
   \end{tabular}
   }$\,, \quad$   \\
\end{center}
\begin{center}
\tighten
   $\Epsilon^{123} \sim{}$
   ${\it 1}\Cdot{}\!\!\!$
   {\footnotesize
   \tabcolsep=1mm
   \renewcommand{\arraystretch}{0.95}
   \begin{tabular}{|c|c|}
     \hline
                  &\ml{\openone} \\
     \hline
     \Pl{\openone}&              \\
     \hline
   \end{tabular}
   }$\,, \quad$
   $\Epsilon^{124} \sim{}$
   ${\it 1}\Cdot{}\!\!\!$
   {\footnotesize
   \tabcolsep=1mm
   \renewcommand{\arraystretch}{0.95}
   \begin{tabular}{|c|c|}
     \hline
                  &\Pl{\sigma_3} \\
     \hline
     \Pl{\sigma_3}&              \\
     \hline
   \end{tabular}
   }$\,, \quad$     \\
\end{center}
\begin{center}
\tighten
   $\Epsilon^{234} \sim{}$
   ${\it 1}\Cdot{}\!\!\!$
   {\footnotesize
   \tabcolsep=1mm
   \renewcommand{\arraystretch}{0.95}
   \begin{tabular}{|c|c|}
     \hline
                  &\Pl{\sigma_1} \\
     \hline
     \Pl{\sigma_1}&              \\
     \hline
   \end{tabular}
   }$\,, \quad$
   $\Epsilon^{314} \sim{}$
   ${\it 1}\Cdot{}\!\!\!$
   {\footnotesize
   \tabcolsep=1mm
   \renewcommand{\arraystretch}{0.95}
   \begin{tabular}{|c|c|}
     \hline
                  &\Pl{\sigma_2} \\
     \hline
     \Pl{\sigma_2}&              \\
     \hline
   \end{tabular}
   }$\,. \quad$       \\
\end{center}
The matrices of spatial vectors for the conjugate space-time in the
approximate representation coincide with the exact matrices for the
three-dimensional case.

Let us introduce matrices $\gamma_K$ in the correspondence with expressions:
\begin{eqnarray*}
&\Epsilon^0     =           \gamma_0 \,, \qquad
\Epsilon^1     =    i\Cdot \gamma_1 \,, \qquad
\Epsilon^2     =    i\Cdot \gamma_2 \,, &\\
&\Epsilon^3     =    i\Cdot \gamma_3 \,, \qquad
\Epsilon^4     =    i\Cdot \gamma_4 \,, &\\
&\Epsilon^{21}  = \gamma_1 \Cdot \gamma_2 = \gamma_{12}  \,, \qquad
\Epsilon^{13}  = \gamma_3 \Cdot \gamma_1 = \gamma_{31}  \,, &\\
&\Epsilon^{32}  = \gamma_2 \Cdot \gamma_3 = \gamma_{23}  \,, \qquad
\Epsilon^{14}  = \gamma_4 \Cdot \gamma_1 = \gamma_{41}  \,, &\\
&\Epsilon^{42}  = \gamma_2 \Cdot \gamma_4 = \gamma_{24}  \,, \qquad
\Epsilon^{34}  = \gamma_4 \Cdot \gamma_3 = \gamma_{43}  \,, &\\
&\Epsilon^{123} = (-i) \Cdot \gamma_1 \Cdot \gamma_2 \Cdot \gamma_3
= (-i) \Cdot \gamma_{123} \,, &\\
&\Epsilon^{124} = (-i) \Cdot \gamma_1 \Cdot \gamma_2 \Cdot \gamma_4
=    (-i) \Cdot \gamma_{124} \,, &\\
&\Epsilon^{234} = (-i) \Cdot \gamma_2 \Cdot \gamma_3 \Cdot \gamma_4
=    (-i) \Cdot \gamma_{234} \,, &\\
&\Epsilon^{314} = (-i) \Cdot \gamma_3 \Cdot \gamma_1 \Cdot \gamma_4
=    (-i) \Cdot \gamma_{314} \,, &\\
&\Epsilon^{1324} =    \gamma_1 \Cdot \gamma_3 \Cdot \gamma_2\Cdot \gamma_4
= \gamma_{1234} \,.&
\end{eqnarray*}
The matrices $\gamma$ make the full set of Dirac matrices. Thus the Dirac
matrices correspond to the approximate representation of basis vectors of the
conjugate Clifford algebra $\BT{C}_4$ in the conjugate algebra $\BT{C}_3$.
Such a representation will be called {\it the first approximate} one and will be
denoted by
\[
     \Tld{R}_1: \BT{C}_4 \rightarrow \BT{C}_3 \,\{
     \Epsilon^{32}, \Epsilon^{13}, \Epsilon^{21}, \Epsilon^0,
     \Epsilon^1, \Epsilon^2, \Epsilon^3, \Epsilon^{123} \}.
\]

The approximate representation
\[
     R_1: \B{C}_4 \rightarrow \B{C}_3 \,\{
     \varepsilon_{32}, \varepsilon_{13}, \varepsilon_{21}, \varepsilon_0,
     \varepsilon_1, \varepsilon_2, \varepsilon_3, \varepsilon_{123} \}
\]
can be considered by an analogous way.
As a result, we obtain
\begin{center}
\tighten
   $\varepsilon_{0} \sim{}$
   ${\it 1}\Cdot{}\!\!\!$
   {\footnotesize
   \tabcolsep=1mm
   \renewcommand{\arraystretch}{0.95}
   \begin{tabular}{|c|c|}
     \hline
     \Pl{\openone}&              \\
     \hline
                  &\Pl{\openone} \\
     \hline
   \end{tabular}
   }$\,, \quad$
   $\varepsilon_{1324} \sim{}$
   $i \Cdot{}\!\!\!$
   {\footnotesize
   \tabcolsep=1mm
   \renewcommand{\arraystretch}{0.95}
   \begin{tabular}{|c|c|}
     \hline
     \Pl{\openone}&              \\
     \hline
                  &\ml{\openone} \\
     \hline
   \end{tabular}
   }$\,, \quad$
\end{center}
\begin{center}
\tighten
   $\varepsilon_{1} \sim{}$
   $a \Cdot{}\!\!\!$
   {\footnotesize
   \tabcolsep=1mm
   \renewcommand{\arraystretch}{0.95}
   \begin{tabular}{|c|c|}
     \hline
                  &\ml{I} \\
     \hline
     \Pl{I}&              \\
     \hline
   \end{tabular}
   }$\,, \quad$
   $\varepsilon_{2} \sim{}$
   $b \Cdot{}\!\!\!$
   {\footnotesize
   \tabcolsep=1mm
   \renewcommand{\arraystretch}{0.95}
   \begin{tabular}{|c|c|}
     \hline
                  &\ml{I} \\
     \hline
     \Pl{I}&              \\
     \hline
   \end{tabular}
   }$\,, \quad$   \\
\end{center}
\begin{center}
\tighten
   $\varepsilon_{3} \sim{}$
   $i \Cdot{}\!\!\!$
   {\footnotesize
   \tabcolsep=1mm
   \renewcommand{\arraystretch}{0.95}
   \begin{tabular}{|c|c|}
     \hline
                  &\ml{\openone} \\
     \hline
     \Pl{\openone}&              \\
     \hline
   \end{tabular}
   }$\,, \quad$
   $\varepsilon_{4} \sim{}$
   $ i \Cdot{}\!\!\!$
   {\footnotesize
   \tabcolsep=1mm
   \renewcommand{\arraystretch}{0.95}
   \begin{tabular}{|c|c|}
     \hline
                  &\Pl{\openone} \\
     \hline
     \Pl{\openone}&              \\
     \hline
   \end{tabular}
   }$\,, \quad$         \\
\end{center}
\begin{center}
\tighten
   $\varepsilon_{21} \sim{}$
   $(-i) \Cdot{}\!\!\!$
   {\footnotesize
   \tabcolsep=1mm
   \renewcommand{\arraystretch}{0.95}
   \begin{tabular}{|c|c|}
     \hline
     \Pl{\openone}&              \\
     \hline
                  &\Pl{\openone} \\
     \hline
   \end{tabular}
   }$\,, \quad$
   $\varepsilon_{13} \sim{}$
   $ b \Cdot{}\!\!\!$
   {\footnotesize
   \tabcolsep=1mm
   \renewcommand{\arraystretch}{0.95}
   \begin{tabular}{|c|c|}
     \hline
     \Pl{I}&              \\
     \hline
                  &\Pl{I} \\
     \hline
   \end{tabular}
   }$\,, \quad$   \\
\end{center}
\begin{center}
\tighten
   $\varepsilon_{32} \sim{}$
   $ a \Cdot{}\!\!\!$
   {\footnotesize
   \tabcolsep=1mm
   \renewcommand{\arraystretch}{0.95}
   \begin{tabular}{|c|c|}
     \hline
     \Pl{I}&              \\
     \hline
                  &\Pl{I} \\
     \hline
   \end{tabular}
   }$\,, \quad$
   $\varepsilon_{14} \sim{}\!\!$
   $ b \Cdot{}\!\!\!$
   {\footnotesize
   \tabcolsep=1mm
   \renewcommand{\arraystretch}{0.95}
   \begin{tabular}{|c|c|}
     \hline
     \Pl{I}&              \\
     \hline
                  &\ml{I} \\
     \hline
   \end{tabular}
   }$\,, \quad$   \\
\end{center}
\begin{center}
\tighten
   $\varepsilon_{42} \sim{}\!\!$
   $ a \Cdot{}\!\!\!$
   {\footnotesize
   \tabcolsep=1mm
   \renewcommand{\arraystretch}{0.95}
   \begin{tabular}{|c|c|}
     \hline
     \Pl{I}&              \\
     \hline
                  &\Pl{I} \\
     \hline
   \end{tabular}
   }$\,, \quad$
   $\varepsilon_{34} \sim{}$
   ${\it 1}\Cdot{}\!\!\!$
   {\footnotesize
   \tabcolsep=1mm
   \renewcommand{\arraystretch}{0.95}
   \begin{tabular}{|c|c|}
     \hline
     \Pl{\openone}&              \\
     \hline
                  &\Pl{\openone} \\
     \hline
   \end{tabular}
   }$\,, \quad$   \\
\end{center}
\begin{center}
\tighten
   $\varepsilon_{123} \sim{}\!\!$
   ${\it 1}\Cdot{}\!\!\!$
   {\footnotesize
   \tabcolsep=1mm
   \renewcommand{\arraystretch}{0.95}
   \begin{tabular}{|c|c|}
     \hline
      \Pl{\openone}&              \\
     \hline
                   &\Pl{\openone} \\
     \hline
   \end{tabular}
   }$\,, \quad$
   $\varepsilon_{124} \sim{}\!\!$
   ${\it 1}\Cdot{}\!\!\!$
   {\footnotesize
   \tabcolsep=1mm
   \renewcommand{\arraystretch}{0.95}
   \begin{tabular}{|c|c|}
     \hline
                  &\Pl{\openone} \\
     \hline
     \ml{\openone}&              \\
     \hline
   \end{tabular}
   }$\,, \quad$     \\
\end{center}
\begin{center}
\tighten
   $\varepsilon_{234} \sim{}\!\!$
   $ b \Cdot{}\!\!\!$
   {\footnotesize
   \tabcolsep=1mm
   \renewcommand{\arraystretch}{0.95}
   \begin{tabular}{|c|c|}
     \hline
                  &\ml{I} \\
     \hline
     \Pl{I}&              \\
     \hline
   \end{tabular}
   }$\,, \quad$
   $\varepsilon_{314} \sim{}\!\!$
   $ a \Cdot{}\!\!\!$
   {\footnotesize
   \tabcolsep=1mm
   \renewcommand{\arraystretch}{0.95}
   \begin{tabular}{|c|c|}
     \hline
             &\ml{I} \\
     \hline
     \Pl{I}  &              \\
     \hline
   \end{tabular}
   }$\,. \quad$       \\
\end{center}

\subsubsection{Second approximate representation}

Consider the representation
\[
     \Tld{R}_2: \BT{C}_4 \rightarrow \BT{C}_2 \,\{
     \Epsilon^{32}, \Epsilon^{13}, \Epsilon^{21}, \Epsilon^0 \}
\]
which will be called {\it the second approximate representation}.  For this
purpose let's assume that, by calculating the parastrophic matrices by
\eref{F13}, the basis vectors with indices $(42,14,1324,34)$,
$(134,234,4,124)$, and $(1,2,3,123)$ are replaced by the basis vectors with
indices $(32,13,21,0)$, respectively.  Then the dimensionality of matrices of
basis vectors of $\BT{C}_4$ is reduced by half with respect to the first
approximate representation and is equal to $4\times 4$ for the real
representation, $2\times 2$ for the complex representation, and $1\times 1$
for the quaternion representation.

For the regular representation of basis vectors of the conjugate space-time in
the algebra $\BT{C}_2$ we have
\begin{flushleft}
\tighten
        \renewcommand{\arraystretch}{0.95}
        \rule{0pt}{11.2mm}
        $\Epsilon^1 \sim{}\!$
        \raisebox{2.85mm}{\footnotesize
        \tabcolsep=1mm
        \renewcommand{\arraystretch}{0}%
        \begin{tabular}{c|cc|cc|}
          \multicolumn{1}{c}{}                &
          \multicolumn{1}{c}{} &
          \multicolumn{1}{c}{\smash{\ii{\it 2}}} &
          \multicolumn{1}{c}{} &
          \multicolumn{1}{c}{\smash{\ii{\it 123}}} \\
          \multicolumn{5}{c}{\rule{0pt}{1.3mm}}\\
          \multicolumn{1}{c}{}                &
          \multicolumn{1}{c}{\smash{\ii{\it 1}}} &
          \multicolumn{1}{c}{} &
          \multicolumn{1}{c}{\smash{\ii{\it 3}}} &
          \multicolumn{1}{c}{} \\
          \multicolumn{5}{c}{\rule{0pt}{2.0mm}}\\
          \multicolumn{1}{c}{}                &
          \multicolumn{1}{c}{} &
          \multicolumn{1}{c}{\smash{\ii{13}}} &
          \multicolumn{1}{c}{} &
          \multicolumn{1}{c}{\smash{\ii{0}}} \\
          \multicolumn{5}{c}{\rule{0pt}{1.3mm}}\\
          \multicolumn{1}{c}{}                &
          \multicolumn{1}{c}{\smash{\ii{32}}} &
          \multicolumn{1}{c}{} &
          \multicolumn{1}{c}{\smash{\ii{21}}} &
          \multicolumn{1}{c}{} \\
          \multicolumn{5}{c}{\rule{0pt}{0.7mm}}\\
          \cline{2-5}
          \rule[-1mm]{0pt}{\LENl}\ib{32} &    &   &   &\m1 \\
          \rule[-1mm]{0pt}{\LENl}\ib{13} &    &   &\p1&    \\
          \cline{2-5}
          \rule[-1mm]{0pt}{\LENl}\ib{21} &    &\m1&   &    \\
          \rule[-1mm]{0pt}{\LENl}\ib{0}  & \p1&   &   &    \\
          \cline{2-5}
        \end{tabular}
        }%
        ${}= (-i) \Cdot {}$
        {%
        \tabcolsep=1mm
        \begin{tabular}{|c|c|}
          \hline
               &\Pl 1 \\
          \hline
          \Pl 1&      \\
          \hline
        \end{tabular}
        }%
        ${}= (-i) \Cdot \sigma^1$
\end{flushleft}
\begin{flushleft}
\tighten
        \renewcommand{\arraystretch}{0.95}
        \rule{0pt}{11.2mm}
        $\Epsilon^2 \sim{}\!$
        \raisebox{2.85mm}{\footnotesize
        \tabcolsep=1mm
        \renewcommand{\arraystretch}{0}%
        \begin{tabular}{c|cc|cc|}
          \multicolumn{1}{c}{}                &
          \multicolumn{1}{c}{} &
          \multicolumn{1}{c}{\smash{\ii{\it 2}}} &
          \multicolumn{1}{c}{} &
          \multicolumn{1}{c}{\smash{\ii{\it 123}}} \\
          \multicolumn{5}{c}{\rule{0pt}{1.3mm}}\\
          \multicolumn{1}{c}{}                &
          \multicolumn{1}{c}{\smash{\ii{\it 1}}} &
          \multicolumn{1}{c}{} &
          \multicolumn{1}{c}{\smash{\ii{\it 3}}} &
          \multicolumn{1}{c}{} \\
          \multicolumn{5}{c}{\rule{0pt}{2.0mm}}\\
          \multicolumn{1}{c}{}                &
          \multicolumn{1}{c}{} &
          \multicolumn{1}{c}{\smash{\ii{13}}} &
          \multicolumn{1}{c}{} &
          \multicolumn{1}{c}{\smash{\ii{0}}} \\
          \multicolumn{5}{c}{\rule{0pt}{1.3mm}}\\
          \multicolumn{1}{c}{}                &
          \multicolumn{1}{c}{\smash{\ii{32}}} &
          \multicolumn{1}{c}{} &
          \multicolumn{1}{c}{\smash{\ii{21}}} &
          \multicolumn{1}{c}{} \\
          \multicolumn{5}{c}{\rule{0pt}{0.7mm}}\\
          \cline{2-5}
          \rule[-1mm]{0pt}{\LENl}\ib{32} &    &   &\m1&    \\
          \rule[-1mm]{0pt}{\LENl}\ib{13} &    &   &   &\m1 \\
          \cline{2-5}
          \rule[-1mm]{0pt}{\LENl}\ib{21} & \p1&   &   &    \\
          \rule[-1mm]{0pt}{\LENl}\ib{0}  &    &\p1&   &    \\
          \cline{2-5}
        \end{tabular}
        }%
        ${}= (-i) \Cdot {}$
        {%
        \tabcolsep=1mm
        \begin{tabular}{|c|c|}
          \hline
               &\ml i\\
          \hline
          \Pl i&    \\
          \hline
        \end{tabular}
        }%
        ${}= (-i) \Cdot \sigma^2$
\end{flushleft}
\begin{flushleft}
\tighten
        \renewcommand{\arraystretch}{0.95}
        \rule{0pt}{11.2mm}
        $\Epsilon^3 \sim{}\!$
        \raisebox{2.85mm}{\footnotesize
        \tabcolsep=1mm
        \renewcommand{\arraystretch}{0}%
        \begin{tabular}{c|cc|cc|}
          \multicolumn{1}{c}{}                &
          \multicolumn{1}{c}{} &
          \multicolumn{1}{c}{\smash{\ii{\it 2}}} &
          \multicolumn{1}{c}{} &
          \multicolumn{1}{c}{\smash{\ii{\it 123}}} \\
          \multicolumn{5}{c}{\rule{0pt}{1.3mm}}\\
          \multicolumn{1}{c}{}                &
          \multicolumn{1}{c}{\smash{\ii{\it 1}}} &
          \multicolumn{1}{c}{} &
          \multicolumn{1}{c}{\smash{\ii{\it 3}}} &
          \multicolumn{1}{c}{} \\
          \multicolumn{5}{c}{\rule{0pt}{2.0mm}}\\
          \multicolumn{1}{c}{}                &
          \multicolumn{1}{c}{} &
          \multicolumn{1}{c}{\smash{\ii{13}}} &
          \multicolumn{1}{c}{} &
          \multicolumn{1}{c}{\smash{\ii{0}}} \\
          \multicolumn{5}{c}{\rule{0pt}{1.3mm}}\\
          \multicolumn{1}{c}{}                &
          \multicolumn{1}{c}{\smash{\ii{32}}} &
          \multicolumn{1}{c}{} &
          \multicolumn{1}{c}{\smash{\ii{21}}} &
          \multicolumn{1}{c}{} \\
          \multicolumn{5}{c}{\rule{0pt}{0.7mm}}\\
          \cline{2-5}
          \rule[-1mm]{0pt}{\LENl}\ib{32} &    &\p1&   &    \\
          \rule[-1mm]{0pt}{\LENl}\ib{13} & \m1&   &   &    \\
          \cline{2-5}
          \rule[-1mm]{0pt}{\LENl}\ib{21} &    &   &   &\m1 \\
          \rule[-1mm]{0pt}{\LENl}\ib{0}  &    &   &\p1&    \\
          \cline{2-5}
        \end{tabular}
        }%
        ${}= (-i) \Cdot{}$
        {%
        \tabcolsep=1mm
        \begin{tabular}{|c|c|}
          \hline
          \ml 1&      \\
          \hline
               &\Pl 1 \\
          \hline
        \end{tabular}
        }%
        ${}= (-i) \Cdot \sigma^3$
\end{flushleft}
\begin{flushleft}
\tighten
        \renewcommand{\arraystretch}{0.95}
        \rule{0pt}{11.2mm}
        $\Epsilon^4 \sim{}\!(-1) \Cdot $
        \raisebox{2.85mm}{\footnotesize
        \tabcolsep=1mm
        \renewcommand{\arraystretch}{0}%
        \begin{tabular}{c|cc|cc|}
          \multicolumn{1}{c}{}                &
          \multicolumn{1}{c}{} &
          \multicolumn{1}{c}{\smash{\ii{\it 234}}} &
          \multicolumn{1}{c}{} &
          \multicolumn{1}{c}{\smash{\ii{\it 124}}} \\
          \multicolumn{5}{c}{\rule{0pt}{1.3mm}}\\
          \multicolumn{1}{c}{}                &
          \multicolumn{1}{c}{\smash{\ii{\it 134}}} &
          \multicolumn{1}{c}{} &
          \multicolumn{1}{c}{\smash{\ii{\it 4}}} &
          \multicolumn{1}{c}{} \\
          \multicolumn{5}{c}{\rule{0pt}{2.0mm}}\\
          \multicolumn{1}{c}{}                &
          \multicolumn{1}{c}{} &
          \multicolumn{1}{c}{\smash{\ii{13}}} &
          \multicolumn{1}{c}{} &
          \multicolumn{1}{c}{\smash{\ii{0}}} \\
          \multicolumn{5}{c}{\rule{0pt}{1.3mm}}\\
          \multicolumn{1}{c}{}                &
          \multicolumn{1}{c}{\smash{\ii{32}}} &
          \multicolumn{1}{c}{} &
          \multicolumn{1}{c}{\smash{\ii{21}}} &
          \multicolumn{1}{c}{} \\
          \multicolumn{5}{c}{\rule{0pt}{0.7mm}}\\
          \cline{2-5}
          \rule[-1mm]{0pt}{\LENl}\ib{32} &    &\m1&   &    \\
          \rule[-1mm]{0pt}{\LENl}\ib{13} &\p1 &   &   &    \\
          \cline{2-5}
          \rule[-1mm]{0pt}{\LENl}\ib{21} &    &   &   &\m1 \\
          \rule[-1mm]{0pt}{\LENl}\ib{0}  &    &   &\p1&    \\
          \cline{2-5}
        \end{tabular}
        }%
        ${}= i \Cdot {}$
        {%
        \tabcolsep=1mm
        \begin{tabular}{|c|c|}
          \hline
          \Pl 1&      \\
          \hline
               &\Pl 1 \\
          \hline
        \end{tabular}
        }%
        ${}= i \Cdot \openone$
\end{flushleft}
Thus the basis vectors of the conjugate geometric space are represented in
algebra $\BT{C}_2$ by the Pauli matrices.  As a result, we obtain for basis
vectors of algebra $\BT{C}_4$
\[
\begin{array}{l*{2}{@{\qquad}l}}
\Epsilon^0     =    {\it 1}\Cdot \openone           \,, &
\Epsilon^{21}  =    i\Cdot \sigma^3    \,, &
\Epsilon^{34}  =    -{\it 1}\Cdot \sigma_3          \,, \\
\Epsilon^1     =    (-i)\Cdot \sigma^1 \,, &
\Epsilon^{13}  =    i\Cdot \sigma^2    \,, &
\Epsilon^{123} =    -{\it 1}\Cdot \openone         \,, \\
\Epsilon^2     =    (-i)\Cdot \sigma^2 \,, &
\Epsilon^{32}  =    i\Cdot \sigma^1    \,, &
\Epsilon^{124} =    {\it 1}\Cdot \sigma_3           \,, \\
\Epsilon^3     =    (-i)\Cdot \sigma^3 \,, &
\Epsilon^{14}  =    -{\it 1}\Cdot \sigma^1          \,, &
\Epsilon^{134} =    -{\it 1}\Cdot \sigma^2          \,, \\
\Epsilon^4     =    i\Cdot \openone    \,, &
\Epsilon^{42}  =    -{\it 1}\Cdot \sigma^2          \,, &
\Epsilon^{234} =    {\it 1}\Cdot \sigma^1           \,, \\
\Epsilon^{1324} =   i\Cdot \openone    \,.
\end{array}
\]

The approximate representation
\[
     R_2: \B{C}_4 \rightarrow \B{C}_2 \,\{
     \varepsilon_{32}, \varepsilon_{13}, \varepsilon_{21}, \varepsilon_0 \}.
\]
can be found by an analogous way.
For example, for the space-time basis vectors we have
\begin{flushleft}
\tighten
        \renewcommand{\arraystretch}{0.95}
        \rule{0pt}{11.2mm}
        $\varepsilon_1 \sim{}\!\!$
        \raisebox{0.95mm}{\footnotesize
        \tabcolsep=1mm
        \renewcommand{\arraystretch}{0}%
        \begin{tabular}{c@{$\;$}c|cc|cc|}
          &\multicolumn{1}{c}{}                &
          \multicolumn{1}{c}{} &
          \multicolumn{1}{c}{\smash{\ii{13}}} &
          \multicolumn{1}{c}{} &
          \multicolumn{1}{c}{\smash{\ii{0}}} \\
          &\multicolumn{5}{c}{\rule{0pt}{1.3mm}}\\
          &\multicolumn{1}{c}{}                &
          \multicolumn{1}{c}{\smash{\ii{32}}} &
          \multicolumn{1}{c}{} &
          \multicolumn{1}{c}{\smash{\ii{21}}} &
          \multicolumn{1}{c}{} \\
          &\multicolumn{5}{c}{\rule{0pt}{0.7mm}}\\
          \cline{3-6}
          \ib{\it 1}   &\rule[-1mm]{0pt}{\LENl}\ib{32} &    &   &   &\p1 \\
          \ib{\it 2}   &\rule[-1mm]{0pt}{\LENl}\ib{13} &    &   &\p1&    \\
          \cline{3-6}
          \ib{\it 3}   &\rule[-1mm]{0pt}{\LENl}\ib{21} &    &\m1&   &    \\
          \ib{\it 123} &\rule[-1mm]{0pt}{\LENl}\ib{0}  & \m1&   &   &    \\
          \cline{3-6}
        \end{tabular}
        }%
        ${}= a \Cdot {}$
        {%
        \tabcolsep=1mm
        \begin{tabular}{|c|c|}
          \hline
               &\Pl 1 \\
          \hline
          \ml 1&      \\
          \hline
        \end{tabular}
        }%
        ${}= a \Cdot I$
\end{flushleft}
\begin{flushleft}
\tighten
        \renewcommand{\arraystretch}{0.95}
        \rule{0pt}{11.2mm}
        $\varepsilon_2 \sim{}\!\!$
        \raisebox{0.95mm}{\footnotesize
        \tabcolsep=1mm
        \renewcommand{\arraystretch}{0}%
        \begin{tabular}{c@{$\;$}c|cc|cc|}
          &\multicolumn{1}{c}{}                &
          \multicolumn{1}{c}{} &
          \multicolumn{1}{c}{\smash{\ii{13}}} &
          \multicolumn{1}{c}{} &
          \multicolumn{1}{c}{\smash{\ii{0}}} \\
          &\multicolumn{5}{c}{\rule{0pt}{1.3mm}}\\
          &\multicolumn{1}{c}{}                &
          \multicolumn{1}{c}{\smash{\ii{32}}} &
          \multicolumn{1}{c}{} &
          \multicolumn{1}{c}{\smash{\ii{21}}} &
          \multicolumn{1}{c}{} \\
          &\multicolumn{5}{c}{\rule{0pt}{0.7mm}}\\
          \cline{3-6}
          \ib{\it 1}   &\rule[-1mm]{0pt}{\LENl}\ib{32} &    &   &\m1&    \\
          \ib{\it 2}   &\rule[-1mm]{0pt}{\LENl}\ib{13} &    &   &   &\p1 \\
          \cline{3-6}
          \ib{\it 3}   &\rule[-1mm]{0pt}{\LENl}\ib{21} & \p1&   &   &    \\
          \ib{\it 123} &\rule[-1mm]{0pt}{\LENl}\ib{0}  &    &\m1&   &    \\
          \cline{3-6}
        \end{tabular}
        }%
        ${}= b \Cdot {}$
        {%
        \tabcolsep=1mm
        \begin{tabular}{|c|c|}
          \hline
                &\Pl 1\\
          \hline
           \ml 1&     \\
          \hline
        \end{tabular}
        }%
        ${}= b \Cdot I$
\end{flushleft}
\begin{flushleft}
\tighten
        \renewcommand{\arraystretch}{0.95}
        \rule{0pt}{11.2mm}
        $\varepsilon_3 \sim{}\!\!$
        \raisebox{0.95mm}{\footnotesize
        \tabcolsep=1mm
        \renewcommand{\arraystretch}{0}%
        \begin{tabular}{c@{$\;$}c|cc|cc|}
          &\multicolumn{1}{c}{}                &
          \multicolumn{1}{c}{} &
          \multicolumn{1}{c}{\smash{\ii{13}}} &
          \multicolumn{1}{c}{} &
          \multicolumn{1}{c}{\smash{\ii{0}}} \\
          &\multicolumn{5}{c}{\rule{0pt}{1.3mm}}\\
          &\multicolumn{1}{c}{}                &
          \multicolumn{1}{c}{\smash{\ii{32}}} &
          \multicolumn{1}{c}{} &
          \multicolumn{1}{c}{\smash{\ii{21}}} &
          \multicolumn{1}{c}{} \\
          &\multicolumn{5}{c}{\rule{0pt}{0.7mm}}\\
          \cline{3-6}
          \ib{\it 1}   &\rule[-1mm]{0pt}{\LENl}\ib{32} &    &\p1&   &    \\
          \ib{\it 2}   &\rule[-1mm]{0pt}{\LENl}\ib{13} & \m1&   &   &    \\
          \cline{3-6}
          \ib{\it 3}   &\rule[-1mm]{0pt}{\LENl}\ib{21} &    &   &   &\p1 \\
          \ib{\it 123} &\rule[-1mm]{0pt}{\LENl}\ib{0}  &    &   &\m1&    \\
          \cline{3-6}
        \end{tabular}
        }%
        ${}= i \Cdot{}$
        {%
        \tabcolsep=1mm
        \begin{tabular}{|c|c|}
          \hline
          \Pl 1&      \\
          \hline
               &\Pl 1 \\
          \hline
        \end{tabular}
        }%
        ${}= i \Cdot \openone$
\end{flushleft}
\begin{flushleft}
\tighten
        \renewcommand{\arraystretch}{0.95}
        \rule{0pt}{11.2mm}
        $\varepsilon_4 \sim{}\!\!$
        \raisebox{0.95mm}{\footnotesize
        \tabcolsep=1mm
        \renewcommand{\arraystretch}{0}%
        \begin{tabular}{c@{$\;$}c|cc|cc|}
          &\multicolumn{1}{c}{}                &
          \multicolumn{1}{c}{} &
          \multicolumn{1}{c}{\smash{\ii{13}}} &
          \multicolumn{1}{c}{} &
          \multicolumn{1}{c}{\smash{\ii{0}}} \\
          &\multicolumn{5}{c}{\rule{0pt}{1.3mm}}\\
          &\multicolumn{1}{c}{}                &
          \multicolumn{1}{c}{\smash{\ii{32}}} &
          \multicolumn{1}{c}{} &
          \multicolumn{1}{c}{\smash{\ii{21}}} &
          \multicolumn{1}{c}{} \\
          &\multicolumn{5}{c}{\rule{0pt}{0.7mm}}\\
          \cline{3-6}
          \ib{\it 134} &\rule[-1mm]{0pt}{\LENl}\ib{32} &    &\p1&   &    \\
          \ib{\it 234} &\rule[-1mm]{0pt}{\LENl}\ib{13} & \m1&   &   &    \\
          \cline{3-6}
          \ib{\it 4}   &\rule[-1mm]{0pt}{\LENl}\ib{21} &    &   &   &\p1 \\
          \ib{\it 124} &\rule[-1mm]{0pt}{\LENl}\ib{0}  &    &   &\m1&    \\
          \cline{3-6}
        \end{tabular}
        }%
        ${}= i \Cdot{}$
        {%
        \tabcolsep=1mm
        \begin{tabular}{|c|c|}
          \hline
          \Pl 1&      \\
          \hline
               &\Pl 1 \\
          \hline
        \end{tabular}
        }%
        ${}= i \Cdot \openone$
\end{flushleft}
As a result, the basis vectors of algebra $\B{C}_4$ are represented as
\[
\begin{array}{l*{2}{@{\qquad}l}}
     \varepsilon_0      =   {\it 1}\Cdot \openone           \,, &
     \varepsilon_{21}   =   i\Cdot \openone    \,, &
     \varepsilon_{34}   =   i\Cdot \openone    \,, \\
     \varepsilon_1      =   a\Cdot I           \,, &
     \varepsilon_{13}   =   b\Cdot I           \,, &
     \varepsilon_{123}  =   {\it 1}\Cdot \openone           \,, \\
     \varepsilon_2      =   b\Cdot I           \,, &
     \varepsilon_{32}   =   a\Cdot I           \,, &
     \varepsilon_{124}  =   {\it 1}\Cdot \openone           \,, \\
     \varepsilon_3      =   i\Cdot \openone    \,, &
     \varepsilon_{14}   =   a\Cdot I           \,, &
     \varepsilon_{134}  =   a\Cdot I           \,, \\
     \varepsilon_4      =   i\Cdot \openone    \,, &
     \varepsilon_{42}   =   b\Cdot I           \,, &
     \varepsilon_{234}  =   b\Cdot I           \,, \\
     \varepsilon_{1324} =   i\Cdot \openone    \,.
\end{array}
\]

\subsubsection{Third approximate representation}

Consider the representation
\[
     \Tld{R}_3: \BT{C}_4 \rightarrow \BT{C}_1 \,\{
     \Epsilon^{21}, \Epsilon^0 \} \,,
\]
which will be called {\it the third approximate representation}.
For this purpose let us assume that the basis vectors with indices $(42,14)$,
$(1324,34)$, $(134,234)$, $(4,124)$, $(1,2)$, $(3,123)$, $(32,13)$ are
replaced by the basis vectors with indices $(21,0)$, respectively. Then the
dimensionality of matrices of basis vectors of $\BT{C}_4$ is reduced by half
with respect to the second approximate representation and is equal to
$2\times 2$ for the real representation, $1\times 1$ for the complex
representation.  As a result, the basis vectors of algebra $\BT{C}_4$ are
represented as
\[
\begin{array}{l*{3}{@{\quad}l}}
     \Epsilon^0     =     {\it 1}  \,, &
     \Epsilon^1     =    -i  \,, &
     \Epsilon^2     =     {\it 1}  \,, &
     \Epsilon^3     =    -i  \,, \\
     \Epsilon^4     =     i  \,, &
     \Epsilon^{21}  =     i  \,, &
     \Epsilon^{13}  =    -{\it 1}  \,, &
     \Epsilon^{32}  =     i  \,, \\
     \Epsilon^{14}  =     {\it 1}  \,, &
     \Epsilon^{42}  =    -i  \,, &
     \Epsilon^{34}  =     {\it 1}  \,, &
     \Epsilon^{123} =    -{\it 1}  \,, \\
     \Epsilon^{124} =     {\it 1}  \,, &
     \Epsilon^{134} =    -i  \,, &
     \Epsilon^{234} =     {\it 1}  \,, &
     \Epsilon^{1324}=     i  \,.
\end{array}
\]

In the approximate representation
\[
     R_3: \B{C}_4 \rightarrow \B{C}_1 \,\{ \varepsilon_{21}, \varepsilon_0 \}
\]
the basis vectors of algebra $\B{C}_4$ are written in the form
\[
\begin{array}{l*{3}{@{\quad}l}}
     \varepsilon_0     =     {\it 1}  \,, &
     \varepsilon_1     =     a  \,, &
     \varepsilon_2     =     b  \,, &
     \varepsilon_3     =     i  \,, \\
     \varepsilon_4     =     i  \,, &
     \varepsilon_{21}  =     i  \,, &
     \varepsilon_{13}  =     b  \,, &
     \varepsilon_{32}  =     a  \,, \\
     \varepsilon_{14}  =     b  \,, &
     \varepsilon_{42}  =     a  \,, &
     \varepsilon_{34}  =     {\it 1}  \,, &
     \varepsilon_{123} =     {\it 1}  \,, \\
     \varepsilon_{124} =     {\it 1}  \,, &
     \varepsilon_{134} =     a  \,, &
     \varepsilon_{234} =     b  \,, &
     \varepsilon_{1324}=     i  \,.
\end{array}
\]

\subsection{Derivation of vectors of generalized space-time. Structure
equations}

The existence of {\it the structure equations} is the important
feature of derivation of algebras. It is connected to derivation of the
multiplication rule for vectors.  We shall consider the structure equations
for the algebra $\B{D}$ being the subalgebra of contravariant universal
algebra $\B{X}$.

Consider vectors $x,x_1,x_2\in \B{D}$ connected by
the multiplication rule:
\begin{equation}
      x=x_1\circ x_2 \,.
\label{F14}
\end{equation}
From \eref{F03} the coordinate form of the multiplication rule follows
\[
     x^K = {\CC^K\!}_{LI}\Cdot ({x_2})^I\Cdot ({x_1})^L \,.
\]
Use the inverse vector $x^{-1}$ for
which the condition \eref{F04} is fulfilled. We obtain for the inverse vector:
\begin{equation}
     x^{-1} = (x_1\circ x_2)^{-1} = (x_2)^{-1}\circ (x_1)^{-1}\,.
\label{F15}
\end{equation}

Let us introduce a {\it differential operator} $\D$ acting on the right
expression. Consider the differential $\D x$.  We shall distinguish
differentials $\D_1$, $\D_2,\ldots$ by index,
the differential of vector $x$ by variation of
vector $x_p$ will be denoted by $\D_p x$.  From \eref{F14} follows
\begin{equation}
     \D x_1 = \D_1 x \circ (x_2)^{-1}\,,\qquad
     \D x_2 = (x_1)^{-1}\circ \D_2 x\,.
\label{F16}
\end{equation}

Let us introduce the second differential $\D_2 \D_1 x$. From \eref{F14}
the second differential is written as
\[
     \D_2 \D_1 x = \D x_1 \circ \D x_2 \,.
\]
Using \eref{F16} and \eref{F15} we get
\begin{equation}
     \D_2 \D_1 x = \D_1 x \circ (x)^{-1} \circ \D_2 x \,.
\label{F17}
\end{equation}
In the neighbourhood of the algebra unit, i.e. when $x = (x)^{-1} =
\varepsilon_0$, the relation  \eref{F17} takes the form
\begin{equation}
     \D_2 \D_1 x=\D_1 x\circ \D_2 x \,.
\label{F18}
\end{equation}
This relation is the {\it structure equation} of algebra $\B{D}$ in the
vector form.
If we substitute in \eref{F18} differentials expressed through basis vectors
$\D x = e_I \cdot \D x^I$ and use the multiplication rule for basis vectors
\eref{F03}, we obtain the structure equations in the coordinate form
\[
     \D_2 \D_1 x^L={\CC^L\!}_{KI}\cdot \D_2 x^I \cdot \D_1 x^K  \,.
\]

The previous considerations are readily generalized to the differential of
$n$-th order $\D_n \D_{n-1}\ldots \D_2 \D_1 x$.  The common structure
equation has the form
\[
     \D_n \D_{n-1}\ldots \D_2 \D_1 x =
     \D_1 x \circ \D_2 x \circ \ldots \circ
     \D_{n-1} x \circ \D_n x \,.
\]

\section{Relativistic quantum mechanics equations and leptons}

\subsection{Generalized action vector. The quantization equations in
differentials}

In this section we shall generalize the notion of action. We suppose that the
action is vector instead of scalar.  A space of {\it action vectors} will
denoted by $\B{SX}$.  We also assume that the space $\B{SX}$ is similarly to
the space $\B{X}$ bearing in mind that the basis vectors of $\B{X}$ can be
accepted as the basis vectors in the space $\B{SX}$.  Thus the action vector
$S\in \B{SX}$ can be written as $S=e_K\cdot S^K$. We endow the coordinates
$S^K$ with the dimensionality $[\text{erg}\times \text{s}]$ in contrast to
dimensionless coondinates of vector of $\B{X}$. Then the scalar component of
this vector, $e_0\Cdot S^0$, is the action in a classical sense.

The set $\B{SX}$ is algebra as well as $\B{X}$ with the same multiplication
rule for basis vectors. The multiplication rule for vectors in algebra
$\B{SX}$ can be written in form
\[
     S = - \frac{1}{S^0}\Cdot S_1 \circ S_2 \,,
\]
where $S, S_1, S_2 \in \B{SX}$, $S^0$ is a constant with the dimensionality
of action used for the agreement of dimensionalities of the right and left
sides of equation.

For algebra $\B{SX}$ as well as for algebra $\B{X}$,
there are structure equations which can be written as follows
\begin{equation}
     \D_2\D_1 S = - \frac{1}{S^0} \Cdot \D_1 S \circ \D_2 S \,.
\label{F19}
\end{equation}
These equations are similar to the equations \eref{F18}. Or in the coordinate
form
\begin{equation}
     \D_2\D_1 S^I =
     - \frac{1}{S^0} \Cdot {\CC^I\!}_{LR} \cdot \D_2 S^R \cdot \D_1 S^L \,.
\label{F20}
\end{equation}
We shall consider the vector $S$ as a function of vector
$x\in \B{X}$: $S=S(x)$.

In the equations \eref{F19} and \eref{F20}, let us introduce the notation
\[
     \psi = \D_1 S
\]
and the notation $\dd$ for the differential $\D_2$.
In the new notations the structure equation takes the form
\[
     \dd \psi = - \frac{1}{S^0}\Cdot \psi\circ \dd S \,.
\]
The vector $\psi$ will be identified with a {\it wave function}.  The structure
equations in the wave function coordinates $\psi^I=\D_1 S^I$:
\[
     \dd \psi^I = - \frac{1}{S^0}\Cdot {\CC^I\!}_{LR} \cdot \dd S^R \cdot
     \psi^L
\]
will be called {\it quantization equations in differentials}.

Express the differential $\dd S$ as
\[
     \dd S = \partial_M S\cdot \dd x^M
\]
and introduce {\it generalized impulses} as
\[
     p_M = - \partial_M S = - e_R\cdot \partial_M S^R =
     e_R\cdot {p^R\!}_M  \,.
\]
Here
\[
     {p^R\!}_M = - \partial_M S^R
\]
are the coordinates of generalized impulses.
From the quantization equations in differentials, it follows that
\begin{equation}
     \partial_M\psi^I(x) =
     \frac{1}{S^0}\Cdot {\CC^I\!}_{LR} \cdot {p^R\!}_M \cdot \psi^L \,.
\label{F21}
\end{equation}
These relations will be called {\it quantum postulates}.

\subsection{Relativistic quantum mechanics equations}

Consider the quantum postulates \eref{F21}. We multiply these relations
by the structure constant ${\CC^{MK}\!}_I$:
\begin{equation}
     {\CC^{MK}\!}_I \cdot \partial_M\psi^I(x) =
     \frac{1}{S^0}\Cdot {\CC^{MK}\!}_I\cdot {\CC^I\!}_{LR}\cdot {p^R\!}_M
     \cdot \psi^L \,.
\label{F22}
\end{equation}
These equations will be called {\it relativistic quantum
mechanics equations in Dirac's form}.

Further we switch from the spaces $\B{X}$ and $\B{SX}$ to them subspaces
$\B{C}_4$ and $\B{SC}_4$. Thus the parastrophic matrices for the Clifford
algebras $\B{C}_4$ and $\BT{C}_4$ will be used as ${C^I\!}_{LR}$ and
${C^{MK\!}}_I$.  In this case the components of wave function, $\psi^I(x)$,
are sixteen real functions.  We set $S^0=\hbar$, the Plank constant. Let also
the wave function $\psi(x)$ depend only on coordinates of the usual
space-time $X$.  Moreover we assume that the generalized impulse ${p^R\!}_M$
has only two components
\begin{eqnarray*}
     {p^0\!}_0 = - \partial_0 S^0 &=& \mc \,,\\
     {p^{34}\!}_0 = - \partial_0 S^{34} &=& \mc \,.
\end{eqnarray*}
Then we obtain
\[
     {\CC^{mK}\!}_I  \Cdot \partial_m\psi^I =
     \frac{\mass \Cdot c}{2\Cdot\hbar} \Cdot ({\delta^K\!}_L + {\CC^K\!}_{L34})
     \Cdot \psi^L \,.
\]
As we shall see later, these relations can be considered as {\it the
relativistic quantum mechanics equations for free leptons}.

Rewrite these equations for the representation of algebra $\B{C}_4$ in
the subalgebra $\B{C}_n$ ($n<4$) over the field of hypernumbers $\B{C}_{4-n}$
in the vector form:
\[
     {\CC^{mK}\!}_I \cdot \partial_m\psi^I\cdot \varepsilon_K  =
     \frac{\mass \Cdot c}{2\Cdot\hbar} \Cdot
     ({\delta^K\!}_L + {\CC^K\!}_{L34}) \Cdot \psi^L \cdot \varepsilon_K \,.
\]
In this expression we represent the basis vector $\varepsilon_K$ as the
product:
\[
     \varepsilon_K = \varepsilon_{k_2} \circ \varepsilon_{k_1}\,,
\]
where $\varepsilon_{k_1}$ are the basis vectors of subalgebra $\B{C}_n$, and
$\varepsilon_{k_2}$ are the basis vectors of subalgebra $\B{C}_{4-n}$. We
next pass from the basis vectors $\varepsilon_{k_2}$ to the basis numbers
$\xi_{k_2}$ in the correspondence with \eref{F06} and to the parastrophic
matrices ${\CC^{mk_1}\!}_{i_1}$ expressed through these numbers. As a result,
we obtain
\begin{equation}
     {\CC^{mk_1}\!}_{i_1}\Cdot \partial_m\Psi^{i_1} =
     \frac{\mass \Cdot c}{2\Cdot\hbar} \Cdot
     ({\delta^{k_1}\!}_{l_1} + {\CC^{k_1}\!}_{{l_1}34})\Cdot \Psi^{l_1} \,,
\label{F22.1}
\end{equation}
where $\Psi^{i_1} = \xi_{i_2}\Cdot \psi^{i_2 i_1}$.

In the complex representation
\begin{eqnarray*}
     \xi_{i_2}         &=& \{{\it 1}, i\} \,,\\
     \varepsilon_{n_1} &=&
     \{\varepsilon_{13}, \varepsilon_0, \varepsilon_{14}, \varepsilon_{34},
     \varepsilon_2, \varepsilon_{123}, \varepsilon_{234}, \varepsilon_{124}\}
\end{eqnarray*}
the components of wave function are complex:
\begin{equation}
\renewcommand{\arraystretch}{1.2}
\begin{array}{l@{}l@{}l@{\qquad}l@{}l@{}l}
     \bpsi^{13}  &{}= i\Cdot \psi^{32}  &{}+ \psi^{13}  \,,&
     \bpsi^0     &{}= i\Cdot \psi^{21}  &{}+ \psi^0     \,,\\
     \bpsi^{14}  &{}= i\Cdot \psi^{42}  &{}+ \psi^{14}  \,,&
     \bpsi^{34}  &{}= i\Cdot \psi^{1324}&{}+ \psi^{34}  \,,\\
     \bpsi^2     &{}= i\Cdot \psi^1     &{}+ \psi^2     \,,&
     \bpsi^{123} &{}= i\Cdot \psi^3     &{}+ \psi^{123} \,,\\
     \bpsi^{234} &{}= i\Cdot \psi^{134} &{}+ \psi^{234} \,,&
     \bpsi^{124} &{}= i\Cdot \psi^4     &{}+ \psi^{124} \,.
\end{array}
\label{F22.5}
\end{equation}

In the quaternion representation
\begin{eqnarray*}
     \xi_{i_2}         &=&
     \{{\it 1}\Cdot \openone, i\Cdot \openone, b\Cdot I, a\Cdot I\} \,,\\
     \varepsilon_{n_1} &=&
     \{\varepsilon_0, \varepsilon_{34}, \varepsilon_{123}, \varepsilon_{124}\}
\end{eqnarray*}
the components of wave function are quaternion:
\begin{equation}
\renewcommand{\arraystretch}{1.2}
\begin{array}{l@{}c*{8}{@{}l}}
     \Psi^0      &{}={}&{}  a\Cdot I\Cdot {}&\psi^{32}  &{}
                          + b\Cdot I\Cdot {}&\psi^{13}  &{}
                          +        i\Cdot {}&\psi^{21}  &{}
                          +                      {}&\psi^0 \,,\\
     \Psi^{34}   &{}={}&{}  a\Cdot I\Cdot {}&\psi^{42}  &{}
                          + b\Cdot I\Cdot {}&\psi^{14}  &{}
                          +        i\Cdot {}&\psi^{1324}&{}
                          +               {}&\psi^{34}     \,,\\
     \Psi^{123}  &{}={}&{}  a\Cdot I\Cdot {}&\psi^1     &{}
                          + b\Cdot I\Cdot {}&\psi^2     &{}
                          +        i\Cdot {}&\psi^3     &{}
                          +                      {}&\psi^{123} \,,\\
     \Psi^{124}  &{}={}&{}  a\Cdot I\Cdot {}&\psi^{134} &{}
                          + b\Cdot I\Cdot {}&\psi^{234} &{}
                          +        i\Cdot {}&\psi^4     &{}
                          +               {}&\psi^{124}\,.
\end{array}
\label{F23}
\end{equation}

Let us write the quantum mechanics equations for the quaternion components:
\begin{center}
{\renewcommand{\arraystretch}{0.95}
\tighten
\tabcolsep=1mm
\begin{tabular}{r}
      $\left( i \, \Cdot
      \hbox{\footnotesize
      \begin{tabular}{|cc|cc|}
        \hline
            &    &    &\p 1\\
            &    &\p 1&    \\
        \hline
            &\p 1&    &     \\
        \p 1&    &    &     \\
        \hline
      \end{tabular}
      }\Cdot \, \partial_4 + i \, \Cdot
      \hbox{\footnotesize
      \begin{tabular}{|cc|cc|}
        \hline
                      &              &\mL{\sigma^a} &                 \\
                      &              &              &\mL{\sigma^a}    \\
        \hline
        \pL{\sigma^a} &              &              &                 \\
                      &\pL{\sigma^a} &              &                 \\
        \hline
      \end{tabular}
      }  \Cdot \, \partial_a \right) \,$
     {\footnotesize
     \begin{tabular}{||l||}
     $\Psi^{\scriptscriptstyle 0  }$   \\
     $\Psi^{\scriptscriptstyle 34 }$   \\
     $\Psi^{\scriptscriptstyle 123}$   \\
     $\Psi^{\scriptscriptstyle 124}$   \\
     \end{tabular}} \\[7mm]
     ${}= \frac{\displaystyle \mass \Cdot c}{\displaystyle 2\Cdot\hbar}
     \, \Cdot $
      {\footnotesize
      \begin{tabular}{|cc|cc|}
        \hline
        \p 1&\p 1&    &    \\
        \p 1&\p 1&    &    \\
        \hline
            &    &\p 1&\p 1 \\
            &    &\p 1&\p 1 \\
        \hline
      \end{tabular}
      }$ \Cdot $
      {\footnotesize
     \begin{tabular}{||l||}
     $\Psi^{\scriptscriptstyle 0  }$   \\
     $\Psi^{\scriptscriptstyle 34 }$   \\
     $\Psi^{\scriptscriptstyle 123}$   \\
     $\Psi^{\scriptscriptstyle 124}$   \\
     \end{tabular}}
\end{tabular}
}%
\end{center}
Or
\begin{equation}
\renewcommand{\arraystretch}{1.8}
\begin{array}{l*{3}{@{}l}}
    i\Cdot \partial_4\Psi^{124}  &{}- i\Cdot\sigma^a\Cdot\partial_a \Psi^{123}
    &{}= \displaystyle \frac{\mass \Cdot c}{2\Cdot\hbar}
    \Cdot (\Psi^0     &{}+ \Psi^{34})                           \,,\\
    i\Cdot \partial_4 \Psi^{123} &{}- i\Cdot\sigma^a\Cdot\partial_a \Psi^{124}
    &{}= \displaystyle \frac{\mass \Cdot c}{2\Cdot\hbar}
    \Cdot (\Psi^0     &{}+ \Psi^{34})                           \,,\\
    i\Cdot \partial_4\Psi^{34}   &{}+ i\Cdot\sigma^a\Cdot\partial_a \Psi^0
    &{}= \displaystyle \frac{\mass \Cdot c}{2\Cdot\hbar}
    \Cdot (\Psi^{123} &{}+ \Psi^{124})                          \,,\\
    i\Cdot \partial_4 \Psi^0     &{}+ i\Cdot\sigma^a\Cdot\partial_a \Psi^{34}
    &{}= \displaystyle \frac{\mass \Cdot c}{2\Cdot\hbar}
    \Cdot (\Psi^{123} &{}+ \Psi^{124}) \,.
\end{array}
\label{F24}
\end{equation}
We  transform these equations as follows. Add the
first equation with the second one and the third one with the fourth one:
\begin{eqnarray}
    i\Cdot \partial_4 \varphi_2
    {}- i\Cdot \sigma^a\Cdot \partial_a \varphi_2
    &=& \frac{\mass\Cdot c}{\hbar} \Cdot \varphi_1 \,,\NN\\
    i\Cdot \partial_4 \varphi_1
    {}+ i\Cdot \sigma^a \Cdot \partial_a \varphi_1
    &=& \frac{\mass\Cdot c}{\hbar} \Cdot \varphi_2 \,,
\label{F25}
\end{eqnarray}
where $\varphi_1=\Psi^0+\Psi^{34}$ and $\varphi_2=\Psi^{123}+\Psi^{124}$.
Then we subtract the third equation from the fourth one, and the first
one from the second one:
\begin{eqnarray*}
     i\Cdot \partial_4\chi_2
     &&{} - i\Cdot \sigma^a \Cdot \partial_a \chi_2 = 0 \,,\\
     i\Cdot \partial_4 \chi_1
     &&{} + i\Cdot \sigma^a \Cdot \partial_a \chi_1 = 0 \,,
\end{eqnarray*}
where $\chi_1=\Psi^0-\Psi^{34}$ and $\chi_2=\Psi^{123}-\Psi^{124}$.
Thus the system of four equations can be transformed to two independent
systems of two equations.

\subsection{Special Cases}

\subsubsection{Dirac theory}

The generalization considered can be reduced to the Dirac theory in the
following way.
\begin{enumerate}
\item
The wave function is represented as the vector of the subalgebra $\B{SC}_3$
constructed on the basis vectors
\[
      \varepsilon_{32}\,, \varepsilon_{13}\,,
      \varepsilon_{21}\,, \varepsilon_0\,,
      \varepsilon_1\,, \varepsilon_2\,,
      \varepsilon_3\,, \varepsilon_{123} \,.
\]
For this the components of wave function
\begin{eqnarray*}
     &&\Psi^{34}(\psi^{42}, \psi^{14}, \psi^{1324}, \psi^{34}) \\
     &&\qquad{}=
     \Psi^{124} (\psi^{134}, \psi^{234}, \psi^4, \psi^{124}) = 0 \,,
\end{eqnarray*}
\begin{eqnarray*}
     \chi_1&=&
     \varphi_1=\Psi^0(\psi^{32}, \psi^{13}, \psi^{21}, \psi^{0})    \,,\\
     \chi_2&=&
     \varphi_2=\Psi^{123}(\psi^{1}, \psi^{2}, \psi^{3}, \psi^{123}) \,.
\end{eqnarray*}

\item
The first approximate representation $\Tld{R}_1$ is used for the basis
vectors of $\BT{C}_4$ (see Section~\ref{Approx_Representation}).

\end{enumerate}
Then the equation system \eref{F22.1} is reduced to the system \eref{F25}
equivalent the Dirac equations.

\subsubsection{Pauli theory}

To obtain the Pauli theory as a special case it is necessary
1) to consider the wave function as the vector of the subalgebra $\B{SC}_2$,
constructed on the basis vectors
\[
      \varepsilon_{32}\,, \varepsilon_{13}\,,
      \varepsilon_{21}\,, \varepsilon_0 \,;
\]
2) to use the second approximate representation $\Tld{R}_2$
for the basis vectors of $\BT{C}_4$.
Then the basis vectors of the conjugate space-time will have the form
\[
      \Epsilon^a = (-i)\Cdot \sigma^a \quad (a=1,2,3)  \,,\qquad
      \Epsilon^4 = i\Cdot \openone       \,.
\]
The equation \eref{F25} can be rewritten as follows
\begin{eqnarray*}
    (\Epsilon^4\Cdot \partial_4 + \Epsilon^a\Cdot \partial_a)&\Psi^{123}&
    {}= \frac{\mass\Cdot c}{\hbar} \Cdot \Psi^0     \,,\\
    (\Epsilon^4\Cdot \partial_4 -\Epsilon^a\Cdot \partial_a) &\Psi^0&
    {}= \frac{\mass\Cdot c}{\hbar} \Cdot \Psi^{123} \,,
\end{eqnarray*}
where $\Psi^0$ and $\Psi^{123}$ are the quaternion functions defined by
\eref{F23}.
Eliminating the quaternion $\Psi^{123}$ from this system
we obtain the equation
\[
    (\Epsilon^4\Cdot \partial_4
    + \Epsilon^a\Cdot \partial_a)\circ
    (\Epsilon^4\Cdot \partial_4
    -\Epsilon^a\Cdot \partial_a)\Psi^0
    = \frac{\mass^2\Cdot c^2}{\hbar^2} \Cdot \Psi^0 \,,
\]
which is reduced to the Klein-Gordon equation with respect to
the quaternion $\Psi^0$:
\[
    (\Epsilon^4\circ\Epsilon^4\Cdot\partial^2_4 -
    \Epsilon^a\circ\Epsilon^b\Cdot\partial_a\partial_b)\Psi^0 =
    \frac{\mass^2\Cdot c^2}{\hbar^2} \Cdot\Psi^0 \,.
\]
This equation is equivalent to the Klein-Gordon equation for two complex
functions
\[
     \Psi^0 =
     \begin{array}{||c@{}c||}
         i \Cdot \psi^{32} &{}+ \psi^{13}\\
         i \Cdot \psi^{21} &{}+ \psi^0
     \end{array} \,.
\]
In the non-relativistic approximation, it is reduced to the Pauli equation.

\subsubsection{Schr\"odinger theory}

To obtain the Schr\"odinger theory as a special case it is necessary
1) to consider the wave function as the vector of the subalgebra $\B{SC}_1$,
constructed on the basis vectors
\[
      \varepsilon_{21}\,, \varepsilon_0 \,;
\]
2) to use the third approximate representation $\Tld{R}_3$
for the basis vectors of $\BT{C}_4$.
Then the basis vectors of the conjugate space-time will have the form
\[
      \Epsilon^1 = -i   \,,\quad
      \Epsilon^2 =  {\it 1}   \,,\quad
      \Epsilon^3 = -i   \,,\quad
      \Epsilon^4 =  i   \,.
\]
The equation \eref{F25} can be rewritten as follows
\begin{eqnarray*}
    \Epsilon^4 \partial_4\bpsi^2
          + \Epsilon^1\partial_1\bpsi^{123}
          -\Epsilon^2 \partial_2\bpsi^{123}
          - \Epsilon^3 \partial_3\bpsi^2
    &=& \frac{\mass\Cdot c}{\hbar} \bpsi^{13}       \,,\\
          \Epsilon^4 \partial_4\bpsi^{123}
          + \Epsilon^1 \partial_1\bpsi^2
          +\Epsilon^2 \partial_2\bpsi^2
          + \Epsilon^3 \partial_3\bpsi^{123}
    &=& \frac{\mass\Cdot c}{\hbar}  \bpsi^0         \,,\\
          \Epsilon^4 \partial_4\bpsi^{13}
          - \Epsilon^1 \partial_1\bpsi^0
          +\Epsilon^2 \partial_2\bpsi^0
          + \Epsilon^3 \partial_3\bpsi^{13}
    &=& \frac{\mass\Cdot c}{\hbar} \bpsi^2          \,,\\
          \Epsilon^4 \partial_4\bpsi^0
          - \Epsilon^1 \partial_1\bpsi^{13}
          -\Epsilon^2 \partial_2\bpsi^{13}
          - \Epsilon^3 \partial_3\bpsi^0
    &=& \frac{\mass\Cdot c}{\hbar} \bpsi^{123}      \,,\\
\end{eqnarray*}
where $\bpsi^0$, $\bpsi^{123}$, $\bpsi^2$, $\bpsi^{13}$ are complex functions
defined by \eref{F22.5}.
If we eliminate from the second equation of this system
the complex functions $\bpsi^{123}$ and $\bpsi^2$ by means of the third
and fourth equations, we obtain the equation with
respect to $\bpsi^0$
\begin{eqnarray*}
    &&(\Epsilon^4\circ\Epsilon^4\Cdot \partial^2_4 -
    \Epsilon^1\circ\Epsilon^1\Cdot \partial^2_1 \\
    &&\qquad
    {}+
    \Epsilon^2\circ\Epsilon^2\Cdot \partial^2_2 -
    \Epsilon^3\circ\Epsilon^3\Cdot \partial^2_3)\bpsi^0
    {}=
    \frac{\mass^2\Cdot c^2}{\hbar^2} \Cdot\bpsi^0 \,,
\end{eqnarray*}
which is equivalent to the Klein-Gordon equation and, in a non-relativistic
approximation, is reduced to the Schr\"odinger equations.

\subsection{Symmetries of wave function components and leptons}

\label{Symmetries}

In this section we give the interpretation of the components
$\Psi^0+\Psi^{34}$, $\Psi^{123}+\Psi^{124}$, $\Psi^{123}-\Psi^{124}$,
$\Psi^0-\Psi^{34}$ of wave function as the wave functions of different
particles.

Our interpretation is based on the following reasons:
\begin{enumerate}

\item
The relativistic quantum mechanics equations obtained can be presented as two
systems of equations, each of them applies to the two-component wave function.

\item
The independence of the specified two systems of equations from each other
allows to refer these systems to different particles.
One of these particles is massive but the other is massless.

\item
With the passage from the generalized equations \eref{F24} to the Dirac
equations, when the component $\Psi^{34}$ and $\Psi^{124}$ became equal to
zero, the component $\Psi^0$ passes in a {\it left} component of the Dirac
wave function, and the component $\Psi^{123}$ passes in a {\it right}
component of the Dirac wave function.

\end{enumerate}

The specified circumstances allow to present facts as follows. The
relativistic quantum mechanics equations concern two
particles whose wave functions have two components.
These particles are leptons of the same generation, i.e.
electron and its neutrino $\{e, \nu_e\}$, muon and its neutrino
$\{\mu, \nu_{\mu}\}$, $\tau$-lepton and its neutrino $\{\tau, \nu_{\tau}\}$.
In our case the neutrino is considered as two-component particle. However
the left neutrino is only observed. This fact can be explained by that
interactions involving the right neutrino is significantly weaker than those
involving the left neutrino.  In our following paper \cite{art2} it will be
shown that such a difference between the left and right neutrinos exists.

In order to answer the question of how wave functions (and quantum mechanics
equations) are distinguished for particles of different generations, we
remind of the feature of the complex representation of the generalized
space-time $\B{C}_4$ (or the action space $\B{SC}_4$).  For this
representation the algebra $\B{C}_4$ is considered as the product
$\B{C}_3\times\B{C}_1$, and any vector of $\varepsilon_{21}$,
$\varepsilon_{13}$, $\varepsilon_{32}$ can be taken as the {\it basic}
basis vector of algebra $\B{C}_1$ (see Section
\ref{Representation_features}).  The first case of basic basis vector
can be put into correspondence with the wave functions of particles of
the first generation $\{e, \nu_e\}$, the second case can be put into
correspondence with the wave functions of particles of the second
generation $\{\mu, \nu_{\mu}\}$, the third case can be put into
correspondence with the wave functions of particles of the third generation
$\{\tau, \nu_{\tau}\}$. The components of wave function of one particle, as
it is accepted, will be called {\it right} and {\it left}, and will be
denoted by lower indices $R$ and $L$, respectively.

Thus we set the following correspondence between the components of wave
function and leptons.
The components of wave function $\psi^A$, where
$A$ = 32, 13, 21, 0, 42, 14, 1324, 34, 1, 2, 3, 123, 134, 234, 4, 124, are
separated into four groups:
\begin{eqnarray*}
     &&\hbox{the left component of electron}\\
     e_L &=&
     \Psi^0     \hbox{\footnotesize $ (\psi^{32}, \psi^{13}, \psi^{21}, \psi^0)$} +
     \Psi^{34}  \hbox{\footnotesize $ (\psi^{42}, \psi^{14}, \psi^{1324}, \psi^{34})$}\,,\\
     &&\hbox{the right component of electron}\\
     e_R &=&
     \Psi^{123} \hbox{\footnotesize $ (\psi^1, \psi^2, \psi^3, \psi^{123})$} +
     \Psi^{124} \hbox{\footnotesize $ (\psi^{134}, \psi^{234}, \psi^4, \psi^{124})$}\,,\\
     &&\hbox{the left component  of $e$-neutrino}\\
     \nu_{eL} &=&
     \Psi^{123} \hbox{\footnotesize $ (\psi^1, \psi^2, \psi^3, \psi^{123})$} -
     \Psi^{124} \hbox{\footnotesize $ (\psi^{134}, \psi^{234}, \psi^4, \psi^{124})$}\,,\\
     &&\hbox{the right component of $e$-neutrino}\\
     \nu_{eR} &=&
     \Psi^0     \hbox{\footnotesize $ (\psi^{32}, \psi^{13}, \psi^{21}, \psi^0)$} -
     \Psi^{34}  \hbox{\footnotesize $ (\psi^{42}, \psi^{14}, \psi^{1324}, \psi^{34})$}\,.
\end{eqnarray*}
The components of wave function for leptons of the second and third generations
differ from the above ones by the cyclic permutation of spatial indices:
\begin{eqnarray*}
     \hbox{for muon and $\mu$-neutrino} &\quad&
     3\rightarrow 2 \,,\; 2\rightarrow 1 \,,\; 1\rightarrow 3 \,;\\
     \hbox{for $\tau$-lepton and $\tau$-neutrino} &\quad&
     3\rightarrow 1 \,,\; 2\rightarrow 3 \,,\; 1\rightarrow 2 \,.
\end{eqnarray*}

The relativistic quantum mechanics equations, for example,
for leptons of the second generation will have the form:
\begin{eqnarray*}
    j\Cdot \partial_4 \mu_R
    {}- j\Cdot \sigma^a \Cdot \partial_a \mu_R
    &=& \frac{\mass_\mu\Cdot c}{\hbar} \Cdot \mu_L \,,\\
    j\Cdot \partial_4 \mu_L
    {}+ j\Cdot \sigma^a \Cdot \partial_a \mu_L
    &=& \frac{\mass_\mu\Cdot c}{\hbar} \Cdot \mu_R \,,\\
    j\Cdot \partial_4 \nu_{\mu R}
    {}- j\Cdot \sigma^a \Cdot \partial_a \nu_{\mu R}
    &=& 0                                          \,,\\
    j\Cdot \partial_4 \nu_{\mu L}
    {}+ j\Cdot \sigma^a \Cdot \partial_a \nu_{\mu L}
    &=& 0                                          \,,\\
    {p^{24}\!}_0 = {p^0\!}_0
    &=& \frac{\mass_\mu\Cdot c}{2} \,.
\end{eqnarray*}
The difference of masses of electron, muon, and of $\tau$-lepton testifies
probably to an anisotropy of directions $\varepsilon_{21}$,
$\varepsilon_{13}$, $\varepsilon_{32}$ in the generalized space-time
$\B{C}_4$ and the action space $\B{SC}_4$.

We give the following definitions. The space of Clifford algebra
$\B{C}_4$ serves to describe lepton motion and thus will be called {\it a
space of leptons}. Respectively, the action space $\B{SC}_4$ will be called
{\it an action space of leptons}.

\subsection{Conjugate action vector. Quantization equations
in differentials for antileptons}

To derive the quantum mechanics equations  for antiparticles it is
necessary to pass from the space $\B{X}$ to the conjugate space $\BT{X}$
which will be called {\it a space of antiparticles} in this connection.
In addition, it is necessary to pass from the action space $\B{SX}$ to
the conjugate space $\BT{SX}$.

We suppose that space $\BT{SX}$ is similarly to space $\BT{X}$.
That is, vector $\Tld{S}\in \BT{SX}$ can be written as
$\Tld{S}=S_K\cdot E^K$. For algebra $\BT{SX}$,
the structure equation takes place
\[
     \D_2\D_1 \Tld{S} =
     -\frac{1}{S_0}\Cdot \D_2 \Tld{S} \circ \D_1 \Tld{S} \,.
\]
Or in the coordinate form:
\begin{equation}
     \D_2\D_1 S_I = -\frac{1}{S_0}\Cdot \D_2 S_P \cdot
     \D_1 S_L\cdot {\CC^{PL}\!}_I  \,.
\label{F27}
\end{equation}
We also suppose that the vector $\Tld{S}$ is a function of vector
$\tld{x}\in \BT{X}$: $\Tld{S}=\Tld{S}(\tld{x})$.

Let us introduce the notations $\psi=\D_1 S$ and $\dd =\D_2$ in the
equations \eref{F27}. The function $\psi(\tld{x})$ will be called {\it a
wave function of antileptons}.  The structure equations are written for the
components of wave function as
\[
     \dd \psi_I = -\frac{1}{S_0}\Cdot
     \dd S_P \cdot\psi_L \cdot{\CC^{PL}\!}_I  \,.
\]
They will be called {\it quantization equations in differentials for
antiparticles}.

Write the differential $\dd \Tld{S}$ as follows
\[
     \dd \Tld{S} = \partial^M\Tld{S}(\tld{x}) \Cdot \dd x_M
\]
and introduce {\it generalized conjugate impulses} as
\[
     p^M = -\partial^M\Tld{S}(\tld{x}) =
     -\partial^M\Tld{S}_P(\tld{x}) \Cdot E^P =
     {\tld{p}^M\!}_P \cdot E^N \,.
\]
From the quantization equations in differentials, the relations follow
\begin{equation}
     \partial^M\psi_I(\tld{x}) =
     \frac{1}{S_0}\Cdot {\tld{p}^M\!}_P \cdot \psi_L \cdot{\CC^{PL}\!}_I \,,
\label{F28}
\end{equation}
which will be called {\it quantum postulates for antiparticles}.

\subsection{Relativistic quantum mechanics equations for antileptons}

Consider the quantum postulates for antiparticles \eref{F28}.
We multiply these relations by the structure constant ${\CC^I\!}_{KM}$:
\[
     \partial^M\psi_I(\tld{x}) \cdot {\CC^I\!}_{KM}  =
     \frac{1}{S_0}\Cdot {\tld{p}^M\!}_P\cdot
     {\CC^{PL}\!}_I\cdot {\CC^I\!}_{KM}\cdot \psi_L \,.
\]

To consider the quantum mechanics equations for antileptons it is
necessary to pass from the conjugate action space $\BT{SX}$ to its
subspace $\BT{C}_4$, which will be named {\it a space of antileptons}
in this connection. The matrices of basis vectors $\Epsilon^I$ of this
space are presented in Appendix~\ref{A4}.  We set again $S_0=\hbar$.
Let also the wave function $\psi_I(\tld{x})$ depend only on coordinates
of the conjugate space-time $\Tld{X}$. Moreover we assume that the
generalized conjugate impulse ${{\tld{p}^M}\!}_P$ has only two
components
\begin{eqnarray*}
     {\tld{p}^{1324}\!}_{1324} =
     \partial^{1324}S_{1324}(\tld{x}) &=& \mc  \,,\\
     {\tld{p}_{123}\!}^0       =
     \partial^0S_{123}(\tld{x})       &=& \mc  \,.
\end{eqnarray*}
Then we obtain
\[
     {\CC^I\!}_{Km} \Cdot \partial^m\psi_I =
     \frac{\mass \Cdot c}{2\Cdot\hbar} \Cdot
     ({\CC^{1324L}\!}_I \cdot {\CC^I\!}_{K1324} + {\CC^{123L}\!}_K)
     \Cdot \psi_L \,.
\]
These equations will be named {\it relativistic quantum mechanics
equations for antileptons}.

Consider these equations for the representation of algebra $\BT{C}_4$ in the
subalgebra $\BT{C}_n$ ($n<4$) over the field of hypernumbers $\BT{C}_{4-n}$
in the vector form:
\begin{eqnarray*}
     &&{\CC^I\!}_{Km} \cdot \partial^m\psi_I \cdot \Epsilon^K \\
     &&\qquad{}=
     \frac{\mass \Cdot c}{2\Cdot\hbar} \Cdot
     ({\CC^{1324L}\!}_I \cdot {\CC^I\!}_{K1324} + {\CC^{123L}\!}_K)
     \Cdot \psi_L \cdot \Epsilon^K \,.
\end{eqnarray*}
In this expression we represent $\Epsilon^K$ as the product:
\[
     \Epsilon^K = \Epsilon^{k_1} \circ \Epsilon^{k_2}
     = \xi^{k_2}\Cdot \Epsilon^{k_1} \,,
\]
where $\Epsilon^{k_1}$ are the basis vectors of subalgebra $\BT{C}_n$,
$\Epsilon^{k_2}$ are the basis vectors of subalgebra $\BT{C}_{4-n}$, and
$\xi^{k_2}$ are the basis numbers. As a result, we obtain
\[
     {\CC^{i_1}\!}_{k_1m} \Cdot \partial^m\Psi_{i_1} =
     \frac{\mass \Cdot c}{2\Cdot\hbar} \Cdot
     ({\CC^{1324{l_1}}\!}_{i_1} \Cdot {\CC^{i_1}\!}_{{k_1}1324} +
     {\CC^{123{l_1}}\!}_{k_1}) \Cdot \Psi_{l_1} \,,
\]
where $\Psi_{i_1} = \psi_{i_1 i_2}\Cdot \xi^{i_2}$, and the parastrophic
matrices ${\CC^{i_1}\!}_{k_1m}$ are expressed through the basis numbers.

In the complex representation
\begin{eqnarray*}
     \xi^{i_2}      &=& \{{\it 1}, i\} \,,\\
     \Epsilon^{n_1} &=&
     \{\Epsilon^{13}, \Epsilon^0, \Epsilon^{14}, \Epsilon^{34},
     \Epsilon^2, \Epsilon^{123}, \Epsilon^{234}, \Epsilon^{124}\}
\end{eqnarray*}
the components of wave function are complex:
\[
\begin{array}{l@{}l@{}l@{\qquad}l@{}l@{}l}
     \bpsi_{13}  &{}= i\Cdot \psi_{32}  &{}+ \psi_{13}  \,,&
     \bpsi_0     &{}= i\Cdot \psi_{21}  &{}+ \psi_0     \,,\\
     \bpsi_{14}  &{}= i\Cdot \psi_{42}  &{}+ \psi_{14}  \,,&
     \bpsi_{34}  &{}= i\Cdot \psi_{1324}&{}+ \psi_{34}  \,,\\
     \bpsi_2     &{}= i\Cdot \psi_1     &{}+ \psi_2     \,,&
     \bpsi_{123} &{}= i\Cdot \psi_3     &{}+ \psi_{123} \,,\\
     \bpsi_{234} &{}= i\Cdot \psi_{134} &{}+ \psi_{234} \,,&
     \bpsi_{124} &{}= i\Cdot \psi_4     &{}+ \psi_{124} \,.
\end{array}
\]

In the quaternion representation
\begin{eqnarray*}
     \xi^{i_2}      &=&
     \{{\it 1}\Cdot \openone,
     i\Cdot \sigma_1, i\Cdot \sigma_2, i\Cdot \sigma_3\} \,,\\
     \Epsilon^{n_1} &=&
     \{ \Epsilon^0, \Epsilon^{34}, \Epsilon^{123}, \Epsilon^{124}\}
\end{eqnarray*}
the components of wave function are quaternion:
\[
\begin{array}{l@{}c*{8}{@{}l}}
     \Psi_0      &{}={}&{}  i\Cdot \sigma_1\Cdot {}&\psi_{32}  &{}
                          + i\Cdot \sigma_2\Cdot {}&\psi_{13}  &{}
                          + i\Cdot \sigma_3\Cdot {}&\psi_{21}  &{}
                          +                      {}&\psi_0     \,,\\
     \Psi_{34}   &{}={}&{}  i\Cdot \sigma_1\Cdot {}&\psi_{42}  &{}
                          + i\Cdot \sigma_2\Cdot {}&\psi_{14}  &{}
                          + i\Cdot \sigma_3\Cdot {}&\psi_{1324}&{}
                          +                      {}&\psi_{34}  \,,\\
     \Psi_{123}  &{}={}&{}  i\Cdot \sigma_1\Cdot {}&\psi_1     &{}
                          + i\Cdot \sigma_2\Cdot {}&\psi_2     &{}
                          + i\Cdot \sigma_3\Cdot {}&\psi_3     &{}
                          +                      {}&\psi_{123} \,,\\
     \Psi_{124}  &{}={}&{}  i\Cdot \sigma_1\Cdot {}&\psi_{134} &{}
                          + i\Cdot \sigma_2\Cdot {}&\psi_{234} &{}
                          + i\Cdot \sigma_3\Cdot {}&\psi_4     &{}
                          +                      {}&\psi_{124}\,.
\end{array}
\]

The quantum mechanics equations in relation to the quaternion
components can be written as follows
\begin{eqnarray*}
    i\Cdot \partial^4 \Psi_{124}      + (a\Cdot I\Cdot \partial^1 +
    b\Cdot I\Cdot \partial^2 + i\Cdot \partial^3)\Cdot \Psi_{123}
    &=& \frac{\mass \Cdot c}{2\Cdot\hbar} \Cdot \varphi_1 \,,\\
    - i\Cdot \partial^4 \Psi_{123} + (a\Cdot I \Cdot \partial^1 +
    b\Cdot I\Cdot \partial^2 - i\Cdot \partial^3)\Cdot \Psi_{124}
    &=& \frac{\mass \Cdot c}{2\Cdot\hbar} \Cdot \varphi_2 \,,\\
    - i\Cdot \partial^4 \Psi_{34}  - (a\Cdot I\Cdot \partial^1 +
    b\Cdot I\Cdot \partial^2 + i\Cdot \partial^3)\Cdot \Psi_0
    &=& \frac{\mass \Cdot c}{2\Cdot\hbar} \Cdot \varphi_1 \,,\\
    i\Cdot \partial^4   \Psi_0     - (a\Cdot I\Cdot \partial^1 +
    b\Cdot I\Cdot \partial^2 - i\Cdot \partial^3)\Cdot \Psi_{34}
    &=& \frac{\mass \Cdot c}{2\Cdot\hbar} \Cdot \varphi_2 \,,
\end{eqnarray*}
where $\varphi_1=\Psi_{123}-\Psi_0$, and $\varphi_2=\Psi_{124}-\Psi_{34}$.
This equation system is readily reduced to the two systems from two
equations:
\begin{eqnarray*}
    i\Cdot \partial^4\Cdot \varphi_2   + (a\Cdot I\Cdot \partial^1 +
    b\Cdot I\Cdot \partial^2 + i\Cdot \partial^3)\Cdot \varphi_1
    &=& \frac{\mass\Cdot c}{\hbar} \Cdot \varphi_1 \,,\\
    - i\Cdot \partial^4\Cdot \varphi_1 + (a\Cdot I\Cdot \partial^1 +
    b\Cdot I\Cdot \partial^2 - i\Cdot \partial^3)\Cdot \varphi_2
    &=& \frac{\mass\Cdot c}{\hbar} \Cdot \varphi_2 \,,
\end{eqnarray*}
and
\begin{eqnarray*}
    i\Cdot \partial^4\Cdot \chi_2   &&{}+ (a\Cdot I\Cdot \partial^1 +
    b\Cdot I\Cdot \partial^2 + i\Cdot \partial^3)\Cdot \chi_1 = 0 \,,\\
    - i\Cdot \partial^4\Cdot \chi_1 &&{}+ (a\Cdot I\Cdot \partial^1 +
    b\Cdot I\Cdot \partial^2 - i\Cdot \partial^3)\Cdot \chi_2 = 0 \,,
\end{eqnarray*}
where $\chi_1=\Psi_{123}+\Psi_0$ and $\chi_2=\Psi_{124}+\Psi_{34}$.
Therefore these equations concern two antiparticles whose wave functions have
two components.  The one of the antiparticles is massive, and the other is
massless.  We consider that the equations obtained describe antileptons of
the same generation.

Repeating the reasons of Section~\ref{Symmetries} one can set correspondence
between the components of wave function and the antileptons of different
generations.

\subsection{Relativistic quantum mechanics equations for arbitrary action
vector}

The relativistic quantum mechanics equations obtained were
derived from the structure equations \eref{F19} for
the action algebra $\B{SX}$.
However these equations are a special case of the common structure equations
similar to the equations \eref{F17} for the algebra $\B{X}$:
\[
     \D_2\D_1 S = - \D_1 S \circ S^{-1} \circ \D_2 S\,.
\]
Using the coordinate form of action vector we obtain
\[
     e_I\cdot \D_2\D_1 S^I =
     - (e_L \circ e_Q \circ e_R)\Cdot
     \D_2 S^R \Cdot (S^{-1})^Q \cdot \D_1 S^L  \,.
\]
From the expression for the product of basis vectors
\[
     e_L \circ e_Q \circ e_R =
     e_I\cdot {\CC^I\!}_{LP}\cdot {\CC^P\!}_{QR} \,,
\]
it follows that
\[
     \D_2\D_1 S^I = - \D_2 S^R \Cdot (S^{-1})^Q \cdot \D_1 S^L \Cdot
     ({\CC^P\!}_{QR} \cdot {\CC^I\!}_{LP}) \,.
\]
If we introduce the wave function $\psi$ as $\D_1 S$,
we obtain the quantization equations in differentials
\[
     \D_2\psi^I = - ({\CC^I\!}_{LP}\cdot {\CC^P\!}_{QR})\Cdot
     \D_2 S^R \Cdot (S^{-1})^Q \cdot \psi^L \,,
\]
From them the quantum postulates follow
\[
     \partial_M\psi^I = ({\CC^I\!}_{LP}\cdot {\CC^P\!}_{QR})\Cdot
     {p^R\!}_M \Cdot (S^{-1})^Q \cdot \psi^L \,.
\]
If we multiply these relations by the structure constants ${\CC^{MK}\!}_I$
of the conjugate algebra, we get
\begin{equation}
     {\CC^{MK}\!}_I \cdot \partial_M\psi^I =
     {\CC^{MK}\!}_I \Cdot ({\CC^I\!}_{LP}\cdot {\CC^P\!}_{QR})\Cdot
     {p^R\!}_M \Cdot (S^{-1})^Q \cdot \psi^L \,.
\label{F29}
\end{equation}
These relations are {\it the relativistic quantum mechanics equations
for the arbitrary action vector}. In the case when the
relativistic quantum mechanics equations are considered in the neighbourhood
of action vector
\[
   S^Q=S^0=\hbar \,,
\]
from \eref{F29} we obtain the equations \eref{F22}.

\section{Conclusions}

We summarize the more important results found in the previous Sections.
\begin{enumerate}
\item
The action for elementary particles should be considered as vector in the
space of contravariant tensors of all ranks $\B{SX}$.  The action for
elementary antiparticles should be considered as vector in the space of
covariant tensors of all ranks $\BT{SX}$.  The specified sets $\B{SX}$ and
$\BT{SX}$, supplied by tensor multiplication, are algebras.  For leptons,
these algebras are reduced to the Clifford algebras $\B{SC}$ and $\BT{SC}$.

\item
The wave function is identified with the partial differential of action
vector.

\item
The quantum equations are the structure equations typical for
the action vector algebra. Thus, the quantization effect is
the consequence of the algebraic structure of the action vector set.

\item
The Clifford algebra $\BT{SC}$ generalizes the Dirac algebra. For the
Clifford algebra the wave function has four quaternion components. In the
Dirac approximation two of them transform to the left component of the Dirac
wave function, and the other components transform to the right one.  It is
one of reasons why the above four components are interpreted as the
components of wave functions of leptons of the same generation.

\item
Any of the basis vectors $\varepsilon_{21}$, $\varepsilon_{13}$,
$\varepsilon_{32}$ can be taken as the basic direction for the complex
representation of wave functions of leptons of the same generation.
Thus these basis vectors should be put into the correspondence with three
generations of leptons. This assumption allows to obtain the quantum equations
for leptons of three generations.

\item
The transformation from the Clifford algebra to the conjugate Clifford
algebra is nonequivalent to the transformation to complex conjugate vectors
and matrices. For this reason, the quantum mechanics equations for
antileptons in Dirac's approximation differ from the appropriate equations of
the Dirac theory.

\item
Both the known quantum mechanics equations and those derived here can be
applied only for the action vector near to the Plank constant.  In a common
case it is necessary to use the equations \eref{F29}.

\end{enumerate}

\acknowledgments

The author is very grateful to E.~A.~Cherkashina for assistance in
preparing manuscript.

\appendix
\section{Regular representation of Clifford algebras}

In this Appendix we give the Clifford algebra parastrophic matrices
which realize the regular representation of basis vectors:
\[
     \varepsilon_I \sim {\CC^L\!}_{KI} \equiv \CC_I\!\!\left({}^L_K\right)\,.
\]

\subsection{Clifford algebra $\B{C}_3$}
\label{A1}

For the Clifford algebra constructed on the geometric space, the
basis vectors are represented by the following matrices.
%
{\renewcommand{\arraystretch}{0.95}
\tighten
\begin{flushleft}
\tabcolsep=1.5mm
\hspace*{0mm}%

\end{flushleft}
}
By the transformation of matrices ${\CC^L\!}_{KI}$ the special notations are
used for matrix blocks. At first, the blocks 2$\times$2 were denoted as
follows
\begin{center}
\tighten
   ${\it 1} ={}\!$
   {\footnotesize
   \tabcolsep=1mm
   \renewcommand{\arraystretch}{0.95}
   \begin{tabular}{|c|c|}
     \hline
     \p1&    \\
     \hline
        &\p1 \\
     \hline
   \end{tabular}
   }$\,, \quad$
   $a ={}\!$
   {\footnotesize
   \tabcolsep=1mm
   \renewcommand{\arraystretch}{0.95}
   \begin{tabular}{|c|c|}
     \hline
        &\p1 \\
     \hline
     \p1&    \\
     \hline
   \end{tabular}
   }$\,, \quad$
   $b ={}\!$
   {\footnotesize
   \tabcolsep=1mm
   \renewcommand{\arraystretch}{0.95}
   \begin{tabular}{|c|c|}
     \hline
     \m1&    \\
     \hline
        &\p1 \\
     \hline
   \end{tabular}
   }$\,, \quad$
   $i ={}\!$
   {\footnotesize
   \tabcolsep=1mm
   \renewcommand{\arraystretch}{0.95}
   \begin{tabular}{|c|c|}
     \hline
        &\p1 \\
     \hline
     \m1&    \\
     \hline
   \end{tabular}
   }$\,.$%
\end{center}
The algebra of numbers $\{{\it 1},a,b,i\}$ is represented by the
multiplication rules:
\begin{eqnarray*}
&a^2=b^2={\it 1} \,,\qquad i^2={\it -1} \,, \qquad
a \Cdot b = - b \Cdot a = i \,,&\\
&a \Cdot i = i \Cdot a = - b \,,\qquad b \Cdot i = i \Cdot b = - a \,.&
\end{eqnarray*}
Thereupon the blocks 2$\times$2 were denotated as follows
\begin{center}
\tighten
   $\openone ={}\!$
   {\footnotesize
   \tabcolsep=1mm
   \renewcommand{\arraystretch}{0.95}
   \begin{tabular}{|c|c|}
     \hline
     \p1&    \\
     \hline
        &\p1 \\
     \hline
   \end{tabular}
   }$\,, \qquad$
   $I ={}\!$
   {\footnotesize
   \tabcolsep=1mm
   \renewcommand{\arraystretch}{0.95}
   \begin{tabular}{|c|c|}
     \hline
        &\p1 \\
     \hline
     \m1&    \\
     \hline
   \end{tabular}
   }$\,.$
\end{center}

\subsection{Clifford algebra $\B{C}_4$}
\label{A2}

For the Clifford algebra constructed on the space-time, the basis vectors
are represented by the following matrices.
%
{\renewcommand{\arraystretch}{0.95}
\tighten
\begin{flushleft}
\fbox{
\tabcolsep=1.5mm
%
      }%
\end{tabular}
}
\end{flushleft}
}

\section{Regular representation of conjugated Clifford algebras}

In this Appendix we give the conjugated Clifford algebra parastrophic matrices
which realize the regular representation of basis vectors:
\[
     \Epsilon^I \sim {\CC^{IK}\!}_L \equiv \CC^I\!\!\left({}^K_L\right) \,.
\]

\subsection{Clifford algebra $\BT{C}_3$}
\label{A3}

For the conjugated Clifford algebra constructed on the geometric
space, the basis vectors are represented by the following matrices.
%
{\renewcommand{\arraystretch}{0.95}
\tighten
\begin{flushleft}
\tabcolsep=1.5mm
\hspace*{0mm}%
\begin{tabular}{lrr}
   \raisebox{-2.3mm}{$\Epsilon^{0\phantom{21}}\,\sim$}&
   {\footnotesize
   \tabcolsep=1mm
   \begin{tabular}{c|cccc|cccc|}
         \multicolumn{2}{r}{}&
         \ii{13}&&
         \multicolumn{1}{c}{\ii{ 0}}&&
         \ii{ 2}&&
         \multicolumn{1}{c}{\ii{123}}\\
         \vsp{-2mm}
         \multicolumn{1}{l}{}&
         \ii{32}&&\ii{21}&
         \multicolumn{1}{l}{}&
         \ii{ 1}&&\ii{ 3}\\
   \cline{2-9}
   \ib{32} &\p1&   &   &   &   &   &   &    \\
   \ib{13} &   &\p1&   &   &   &   &   &    \\
   \ib{21} &   &   &\p1&   &   &   &   &    \\
   \ib{0}  &   &   &   &\p1&   &   &   &    \\
   \cline{2-9}
   \ib{1}  &   &   &   &   &\p1&   &   &    \\
   \ib{2}  &   &   &   &   &   &\p1&   &    \\
   \ib{3}  &   &   &   &   &   &   &\p1&    \\
   \ib{123}&   &   &   &   &   &   &   &\p1 \\
   \cline{2-9}
   \end{tabular}
   }%
&
\begin{tabular}{r}
   \vsp{0.4mm}
   \raisebox{-2.6mm}{=\hspace*{2\tabcolsep}%
   ${\it 1} \,\Cdot$}%
   {\footnotesize
   \tabcolsep=1mm
   \begin{tabular}{|cc|cc|c}
         \multicolumn{1}{c}{}&
         \multicolumn{1}{c}{\ii{0}}&&
         \multicolumn{1}{c}{\ii{123}}\\
         \vsp{-2mm}
         \multicolumn{1}{c}{\ii{13}}&
         \multicolumn{1}{c}{}&
         \multicolumn{1}{c}{\ii{ 2}}\\
     \cline{1-4}
     \p1&   &   &   &\ib{13} \\
        &\p1&   &   &\ib{0}  \\
     \cline{1-4}
        &   &\p1&   &\ib{2}  \\
        &   &   &\p1&\ib{123}\\
     \cline{1-4}
   \end{tabular}
   }%
   \\
   \vsp{1.2mm}%
   \raisebox{-1.9mm}{=\hspace*{2\tabcolsep}%
   ${\it 1} \,\Cdot$}%
   {%
   \tabcolsep=1mm
   \begin{tabular}{|c|c|c}
     \multicolumn{1}{c}{\ii{0}}&
     \multicolumn{1}{c}{\ii{123}}\\
     \cline{1-2}
     \Pl{\openone}&             &\ib{2}  \\
     \cline{1-2}
                  &\Pl{\openone}&\ib{123}\\
     \cline{1-2}
   \end{tabular}
   }%
\end{tabular}
\end{tabular}
\end{flushleft}

\begin{flushleft}
\tabcolsep=1.5mm
\hspace*{0mm}%
\begin{tabular}{lrr}
   \raisebox{-2.3mm}{$\Epsilon^{1\phantom{23}}\,\sim$}&
   {\footnotesize
   \tabcolsep=1mm
   \begin{tabular}{c|cccc|cccc|}
         \multicolumn{2}{r}{}&
         \ii{13}&&
         \multicolumn{1}{c}{\ii{ 0}}&&
         \ii{ 2}&&
         \multicolumn{1}{c}{\ii{123}}\\
         \vsp{-2mm}
         \multicolumn{1}{l}{}&
         \ii{32}&&\ii{21}&
         \multicolumn{1}{l}{}&
         \ii{ 1}&&\ii{ 3}\\
   \cline{2-9}
   \ib{32} &   &   &   &   &   &   &   &\m1 \\
   \ib{13} &   &   &   &   &   &   &\p1&    \\
   \ib{21} &   &   &   &   &   &\m1&   &    \\
   \ib{0}  &   &   &   &   &\p1&   &   &    \\
   \cline{2-9}
   \ib{1}  &   &   &   &\p1&   &   &   &    \\
   \ib{2}  &   &   &\m1&   &   &   &   &    \\
   \ib{3}  &   &\p1&   &   &   &   &   &    \\
   \ib{123}&\m1&   &   &   &   &   &   &    \\
   \cline{2-9}
   \end{tabular}
   }%
&
\begin{tabular}{r}
   \vsp{5.0mm}
   =\hspace*{2\tabcolsep}%
   $i \,\Cdot$
   {\footnotesize
   \tabcolsep=1mm
   \begin{tabular}{|cc|cc|}
     \hline
        &   &   &\m1 \\
        &   &\m1&    \\
     \hline
        &\p1&   &    \\
     \p1&   &   &    \\
     \hline
   \end{tabular}
   }%
   \\
   \vsp{4.3mm}%
   =\hspace*{2\tabcolsep}%
   $i \,\Cdot$
   {%
   \tabcolsep=1mm
   \begin{tabular}{|c|c|}
     \hline
                     &\mL{\sigma^1} \\
     \hline
     \pL{\sigma^1} &                \\
     \hline
   \end{tabular}
   }%
\end{tabular}
\end{tabular}
\end{flushleft}

\begin{flushleft}
\tabcolsep=1.5mm
\hspace*{0mm}%
\begin{tabular}{lrr}
   \raisebox{-2.3mm}{$\Epsilon^{2\phantom{31}}\,\sim$}&
   {\footnotesize
   \tabcolsep=1mm
   \begin{tabular}{c|cccc|cccc|}
         \multicolumn{2}{r}{}&
         \ii{13}&&
         \multicolumn{1}{c}{\ii{ 0}}&&
         \ii{ 2}&&
         \multicolumn{1}{c}{\ii{123}}\\
         \vsp{-2mm}
         \multicolumn{1}{l}{}&
         \ii{32}&&\ii{21}&
         \multicolumn{1}{l}{}&
         \ii{ 1}&&\ii{ 3}\\
   \cline{2-9}
   \ib{32} &   &   &   &   &   &   &\m1&    \\
   \ib{13} &   &   &   &   &   &   &   &\m1 \\
   \ib{21} &   &   &   &   &\p1&   &   &    \\
   \ib{0}  &   &   &   &   &   &\p1&   &    \\
   \cline{2-9}
   \ib{1}  &   &   &\p1&   &   &   &   &    \\
   \ib{2}  &   &   &   &\p1&   &   &   &    \\
   \ib{3}  &\m1&   &   &   &   &   &   &    \\
   \ib{123}&   &\m1&   &   &   &   &   &    \\
   \cline{2-9}
   \end{tabular}
   }%
&
\begin{tabular}{r}
   \vsp{5.0mm}
   =\hspace*{2\tabcolsep}%
   $i \,\Cdot$
   {\footnotesize
   \tabcolsep=1mm
   \begin{tabular}{|cc|cc|}
     \hline
         &    &    &\p i \\
         &    &\m i&     \\
     \hline
         &\m i&    &     \\
     \p i&    &    &     \\
     \hline
   \end{tabular}
   }%
   \\
   \vsp{4.3mm}%
   =\hspace*{2\tabcolsep}%
   $i \,\Cdot$
   {%
   \tabcolsep=1mm
   \begin{tabular}{|c|c|}
     \hline
                      &\mL{\sigma^2}   \\
     \hline
     \pL{\sigma^2}  &                  \\
     \hline
   \end{tabular}
   }%
\end{tabular}
\end{tabular}
\end{flushleft}

\begin{flushleft}
\tabcolsep=1.5mm
\hspace*{0mm}%
\begin{tabular}{lrr}
   \raisebox{-2.3mm}{$\Epsilon^{3\phantom{21}}\,\sim$}&
   {\footnotesize
   \tabcolsep=1mm
   \begin{tabular}{c|cccc|cccc|}
         \multicolumn{2}{r}{}&
         \ii{13}&&
         \multicolumn{1}{c}{\ii{ 0}}&&
         \ii{ 2}&&
         \multicolumn{1}{c}{\ii{123}}\\
         \vsp{-2mm}
         \multicolumn{1}{l}{}&
         \ii{32}&&\ii{21}&
         \multicolumn{1}{l}{}&
         \ii{ 1}&&\ii{ 3}\\
   \cline{2-9}
   \ib{32} &   &   &   &   &   &\p1&   &    \\
   \ib{13} &   &   &   &   &\m1&   &   &    \\
   \ib{21} &   &   &   &   &   &   &   &\m1 \\
   \ib{0}  &   &   &   &   &   &   &\p1&    \\
   \cline{2-9}
   \ib{1}  &   &\m1&   &   &   &   &   &    \\
   \ib{2}  &\p1&   &   &   &   &   &   &    \\
   \ib{3}  &   &   &   &\p1&   &   &   &    \\
   \ib{123}&   &   &\m1&   &   &   &   &    \\
   \cline{2-9}
   \end{tabular}
   }%
&
\begin{tabular}{r}
   \vsp{5.0mm}
   =\hspace*{2\tabcolsep}%
   $i \,\Cdot$
   {\footnotesize
   \tabcolsep=1mm
   \begin{tabular}{|cc|cc|}
     \hline
        &   &\p1&    \\
        &   &   &\m1 \\
     \hline
     \m1&   &   &    \\
        &\p1&   &    \\
     \hline
   \end{tabular}
   }%
   \\
   \vsp{4.3mm}%
   =\hspace*{2\tabcolsep}%
   $i \,\Cdot$
   {%
   \tabcolsep=1mm
   \begin{tabular}{|c|c|}
     \hline
                      &\mL{\sigma^3}   \\
     \hline
     \pL{\sigma^3}  &                  \\
     \hline
   \end{tabular}
   }%
\end{tabular}
\end{tabular}
\end{flushleft}

\begin{flushleft}
\tabcolsep=1.5mm
\hspace*{0mm}%
\begin{tabular}{lrr}
   \raisebox{-2.3mm}{$\Epsilon^{21\phantom{3}}\,\sim (-1)\Cdot{}\!\!\! $}&
   {\footnotesize
   \tabcolsep=1mm
   \begin{tabular}{c|cccc|cccc|}
         \multicolumn{2}{r}{}&
         \ii{13}&&
         \multicolumn{1}{c}{\ii{ 0}}&&
         \ii{ 2}&&
         \multicolumn{1}{c}{\ii{123}}\\
         \vsp{-2mm}
         \multicolumn{1}{l}{}&
         \ii{32}&&\ii{21}&
         \multicolumn{1}{l}{}&
         \ii{ 1}&&\ii{ 3}\\
   \cline{2-9}
   \ib{32} &   &\p1&   &   &   &   &   &    \\
   \ib{13} &\m1&   &   &   &   &   &   &    \\
   \ib{21} &   &   &   &\m1&   &   &   &    \\
   \ib{0}  &   &   &\p1&   &   &   &   &    \\
   \cline{2-9}
   \ib{1}  &   &   &   &   &   &\p1&   &    \\
   \ib{2}  &   &   &   &   &\m1&   &   &    \\
   \ib{3}  &   &   &   &   &   &   &   &\m1 \\
   \ib{123}&   &   &   &   &   &   &\p1&    \\
   \cline{2-9}
   \end{tabular}
   }%
&
\begin{tabular}{r}
   \vsp{5.0mm}
   =\hspace*{2\tabcolsep}%
   $i \,\Cdot$
   {\footnotesize
   \tabcolsep=1mm
   \begin{tabular}{|cc|cc|}
     \hline
     \m1&   &   &    \\
        &\p1&   &    \\
     \hline
        &   &\m1&    \\
        &   &   &\p1 \\
     \hline
   \end{tabular}
   }%
   \\
   \vsp{4.3mm}%
   =\hspace*{2\tabcolsep}%
   $i \,\Cdot$
   {%
   \tabcolsep=1mm
   \begin{tabular}{|c|c|}
     \hline
     \Pl{\sigma^3}  &                  \\
     \hline
                      &\Pl{\sigma^3}   \\
     \hline
   \end{tabular}
   }%
\end{tabular}
\end{tabular}
\end{flushleft}

\begin{flushleft}
\tabcolsep=1.5mm
\hspace*{0mm}%
\begin{tabular}{lrr}
   \raisebox{-2.3mm}{$\Epsilon^{13\phantom{2}}\,\sim (-1)\Cdot{}\!\!\! $}&
   {\footnotesize
   \tabcolsep=1mm
   \begin{tabular}{c|cccc|cccc|}
         \multicolumn{2}{r}{}&
         \ii{13}&&
         \multicolumn{1}{c}{\ii{ 0}}&&
         \ii{ 2}&&
         \multicolumn{1}{c}{\ii{123}}\\
         \vsp{-2mm}
         \multicolumn{1}{l}{}&
         \ii{32}&&\ii{21}&
         \multicolumn{1}{l}{}&
         \ii{ 1}&&\ii{ 3}\\
   \cline{2-9}
   \ib{32} &   &   &\m1&   &   &   &   &    \\
   \ib{13} &   &   &   &\m1&   &   &   &    \\
   \ib{21} &\p1&   &   &   &   &   &   &    \\
   \ib{0}  &   &\p1&   &   &   &   &   &    \\
   \cline{2-9}
   \ib{1}  &   &   &   &   &   &   &\m1&    \\
   \ib{2}  &   &   &   &   &   &   &   &\m1 \\
   \ib{3}  &   &   &   &   &\p1&   &   &    \\
   \ib{123}&   &   &   &   &   &\p1&   &    \\
   \cline{2-9}
   \end{tabular}
   }%
&
\begin{tabular}{r}
   \vsp{5.0mm}
   =\hspace*{2\tabcolsep}%
   $i \,\Cdot$
   {\footnotesize
   \tabcolsep=1mm
   \begin{tabular}{|cc|cc|}
     \hline
         &\m i&    &     \\
     \p i&    &    &     \\
     \hline
         &    &    &\m i \\
         &    &\p i&     \\
     \hline
   \end{tabular}
   }%
   \\
   \vsp{4.3mm}%
   =\hspace*{2\tabcolsep}%
   $i \,\Cdot$
   {%
   \tabcolsep=1mm
   \begin{tabular}{|c|c|}
     \hline
     \Pl{\sigma^2}  &                  \\
     \hline
                      &\Pl{\sigma^2}   \\
     \hline
   \end{tabular}
   }%
\end{tabular}
\end{tabular}
\end{flushleft}

\begin{flushleft}
\tabcolsep=1.5mm
\hspace*{0mm}%
\begin{tabular}{lrr}
   \raisebox{-2.3mm}{$\Epsilon^{32\phantom{1}}\,\sim (-1)\Cdot{}\!\!\! $}&
   {\footnotesize
   \tabcolsep=1mm
   \begin{tabular}{c|cccc|cccc|}
         \multicolumn{2}{r}{}&
         \ii{13}&&
         \multicolumn{1}{c}{\ii{ 0}}&&
         \ii{ 2}&&
         \multicolumn{1}{c}{\ii{123}}\\
         \vsp{-2mm}
         \multicolumn{1}{l}{}&
         \ii{32}&&\ii{21}&
         \multicolumn{1}{l}{}&
         \ii{ 1}&&\ii{ 3}\\
   \cline{2-9}
   \ib{32} &   &   &   &\m1&   &   &   &    \\
   \ib{13} &   &   &\p1&   &   &   &   &    \\
   \ib{21} &   &\m1&   &   &   &   &   &    \\
   \ib{0}  &\p1&   &   &   &   &   &   &    \\
   \cline{2-9}
   \ib{1}  &   &   &   &   &   &   &   &\m1 \\
   \ib{2}  &   &   &   &   &   &   &\p1&    \\
   \ib{3}  &   &   &   &   &   &\m1&   &    \\
   \ib{123}&   &   &   &   &\p1&   &   &    \\
   \cline{2-9}
   \end{tabular}
   }%
&
\begin{tabular}{r}
   \vsp{5.0mm}
   =\hspace*{2\tabcolsep}%
   $i \,\Cdot$
   {\footnotesize
   \tabcolsep=1mm
   \begin{tabular}{|cc|cc|}
     \hline
        &\p1&   &    \\
     \p1&   &   &    \\
     \hline
        &   &   &\p1 \\
        &   &\p1&    \\
     \hline
   \end{tabular}
   }%
   \\
   \vsp{4.3mm}%
   =\hspace*{2\tabcolsep}%
   $i \,\Cdot$
   {%
   \tabcolsep=1mm
   \begin{tabular}{|c|c|}
     \hline
     \Pl{\sigma^1}  &                  \\
     \hline
                      &\Pl{\sigma^1}   \\
     \hline
   \end{tabular}
   }%
\end{tabular}
\end{tabular}
\end{flushleft}

\begin{flushleft}
\tabcolsep=1.5mm
\hspace*{0mm}%
\begin{tabular}{lrr}
   \raisebox{-2.3mm}{$\Epsilon^{123}\,\sim (-1)\Cdot{}\!\!\! $}&
   {\footnotesize
   \tabcolsep=1mm
   \begin{tabular}{c|cccc|cccc|}
         \multicolumn{2}{r}{}&
         \ii{13}&&
         \multicolumn{1}{c}{\ii{ 0}}&&
         \ii{ 2}&&
         \multicolumn{1}{c}{\ii{123}}\\
         \vsp{-2mm}
         \multicolumn{1}{l}{}&
         \ii{32}&&\ii{21}&
         \multicolumn{1}{l}{}&
         \ii{ 1}&&\ii{ 3}\\
   \cline{2-9}
   \ib{32} &   &   &   &   &\p1&   &   &    \\
   \ib{13} &   &   &   &   &   &\p1&   &    \\
   \ib{21} &   &   &   &   &   &   &\p1&    \\
   \ib{0}  &   &   &   &   &   &   &   &\p1 \\
   \cline{2-9}
   \ib{1}  &\m1&   &   &   &   &   &   &    \\
   \ib{2}  &   &\m1&   &   &   &   &   &    \\
   \ib{3}  &   &   &\m1&   &   &   &   &    \\
   \ib{123}&   &   &   &\m1&   &   &   &    \\
   \cline{2-9}
   \end{tabular}
   }%
&
\begin{tabular}{r}
   \vsp{5.0mm}
   =\hspace*{2\tabcolsep}%
   ${\it 1} \,\Cdot$
   {\footnotesize
   \tabcolsep=1mm
   \begin{tabular}{|cc|cc|}
     \hline
        &   &\m1&    \\
        &   &   &\m1 \\
     \hline
     \p1&   &   &    \\
        &\p1&   &    \\
     \hline
   \end{tabular}
   }%
   \\
   \vsp{4.3mm}%
   =\hspace*{2\tabcolsep}%
   ${\it 1} \,\Cdot$
   {%
   \tabcolsep=1mm
   \begin{tabular}{|c|c|}
     \hline
                  &\ml{\openone}\\
     \hline
     \Pl{\openone}&             \\
     \hline
   \end{tabular}
   }%
\end{tabular}
\end{tabular}
\end{flushleft}
}
By the transformation of matrices ${\CC^{IK}\!}_L$ the special notations are
used for matrix blocks. At first, blocks 2$\times$2 were denoted as follows
\begin{center}
\tighten
   ${\it 1} ={}\!$
   {\footnotesize
   \tabcolsep=1mm
   \renewcommand{\arraystretch}{0.95}
   \begin{tabular}{|c|c|}
     \hline
     \p1&    \\
     \hline
        &\p1 \\
     \hline
   \end{tabular}
   }$\,, \qquad$
   $i ={}\!$
   {\footnotesize
   \tabcolsep=1mm
   \renewcommand{\arraystretch}{0.95}
   \begin{tabular}{|c|c|}
     \hline
        &\p1 \\
     \hline
     \m1&    \\
     \hline
   \end{tabular}
   }$\,.$%
\end{center}
This is tantamount to changing from real matrices 2$\times$2 to complex
numbers.  Thereupon the complex matrices 2$\times$2 were denoted as follows
\begin{center}
\tighten
   $\openone ={}\!$
   {\footnotesize
   \tabcolsep=1mm
   \renewcommand{\arraystretch}{0.95}
   \begin{tabular}{|c|c|}
     \hline
     \p1&    \\
     \hline
        &\p1 \\
     \hline
   \end{tabular}
   }$\,, \quad$
   $\sigma^1 ={}\!$
   {\footnotesize
   \tabcolsep=1mm
   \renewcommand{\arraystretch}{0.95}
   \begin{tabular}{|c|c|}
     \hline
        &\p1 \\
     \hline
     \p1&    \\
     \hline
   \end{tabular}
   }$\,, \quad$
   $\sigma^2 ={}\!$
   {\footnotesize
   \tabcolsep=1mm
   \renewcommand{\arraystretch}{0.95}
   \begin{tabular}{|c|c|}
     \hline
        &\m i \\
     \hline
     \p i&    \\
     \hline
   \end{tabular}
   }$\,, \quad$
   $\sigma^3 ={}\!$
   {\footnotesize
   \tabcolsep=1mm
   \renewcommand{\arraystretch}{0.95}
   \begin{tabular}{|c|c|}
     \hline
     \m1&    \\
     \hline
        &\p1 \\
     \hline
   \end{tabular}
   }$\,.$
\end{center}
The matrices $\sigma^1$, $\sigma^2$, $\sigma^3$ are the {\it Pauli matrices},
with the difference that the opposite sign matrix was denoted as
$\sigma^3$ for reasons of symmetry.

\subsection{Clifford algebra $\BT{C}_4$}
\label{A4}

For the conjugated Clifford algebra constructed on the conjugated space-time,
the basis vectors are represented by the following matrices.
%
{\renewcommand{\arraystretch}{0.95}
\tighten
\begin{flushleft}
\fbox{
\tabcolsep=1.5mm

      }}%
\end{tabular}
}
\end{flushleft}
}


\end{document}